\documentclass[twocolumn,showpacs,showkeys,preprintnumbers,aps]{revtex4}

\usepackage{graphicx}
\usepackage{graphics}
\usepackage{dcolumn}
\usepackage{bm}
\usepackage{epsfig}
\usepackage{longtable}

\newlength\figwidth\figwidth=0.5\textwidth
\renewcommand{\thefootnote}{\fnsymbol{footnote}}

\begin{document}

\title{Spectroscopy of $^{232}$U in the (p,t) reaction: More information on  $0^+$ excitations }

\author{A.~I.~Levon$^{1}$, P.~Alexa$^{2}$,
 G.~Graw$^{3}$, R.~Hertenberger$^{3}$, S.~Pascu$^{4}$,
 P.~G.~Thirolf$^{3}$,  and H.-F.~Wirth$^{3}$ }

\affiliation{$^1$ Institute for Nuclear Research, Academy of Science, Kiev, Ukraine}
\email[Electronic address: ] {levon@kinr.kiev.ua}

\affiliation{$^2$ Institute of Physics and Institute of Clean Technologies,  Technical
University of Ostrava, Czech Republic}

\affiliation{$^3$ Fakult\"at f\"ur Physik, Ludwig-Maximilians-Universit\"at M\"unchen, Garching,
Germany}

\affiliation{$^4$ H.~Hulubei National Institute of Physics and Nuclear
Engineering, Bucharest, Romania}

\begin{abstract}

 The excitation spectra in the deformed nucleus $^{232}$U have been studied by means of
the (p,t) reaction, using the Q3D spectrograph facility at the Munich Tandem accelerator. The
angular distributions of tritons were measured for 162 excitations seen in the triton spectra up
 to 3.25 MeV.   $0^+$ assignments are made for 13 excited states by
 comparison of experimental angular distributions with the calculated
 ones using the CHUCK3 code.
 Assignments up to spin $6^+$  are made for other states. Sequences of  states
 are selected which can be treated as rotational bands. Moments of inertia have been
 derived from these sequences, whose values may be
 considered as evidence of the  two- or one-phonon nature of these $0^+$ excitations.
 Experimental data are compared with  interacting boson model (IBM)  and
 quasiparticle-phonon model (QPM) calculations.

\end{abstract}
\bigskip
\date{\today}


\maketitle

\section{Introduction}

The first observation of multiple excitations with zero angular momentum
transfer in the (p,t) reaction seen in the odd nucleus $^{229}$Pa~\cite{Lev94}
initiated an extensive campaign to study $0^+$ excitations in even-even actinide nuclei.
During the last two decades, many of such investigations have been performed
using the Q3D magnetic spectrograph at the Maier-Leibnitz-Laboratory (MLL) Tandem
accelerator in Garching.
Because of its very high energy resolution, this spectrograph is a unique tool in
particular for the identification of $0^+$ states by measuring the state-selective angular
distributions of triton ejectiles. Subsequent analysis is performed within the
distorted-wave Born approximation (DWBA).
In addition to our studies on the actinide nuclei $^{230}$Th, $^{228}$Th, $^{232}$U,
the neighbouring odd nucleus $^{229}$Pa~\cite{Wir04}, and  most recently on
$^{240}$Pu~\cite{Spi13},
the majority of studies on 0$^+$ excitations was carried out in the regions of rare earth,
transitional and spherical nuclei~\cite{Les02,Mey06,Buc06,Bet09,Ili10,Ber13}.
Most of these studies were limited to measuring the energies
and excitation cross sections of $0^+$ states.
Therefore they provided only the trend of changes in the nuclei contributing to such
excitations in a wide range of deformations: from transitional nuclei (Gd region) to
well-deformed (Yb region), gamma-soft (Pt region) and spherical nuclei (Pb region).
The main result of these studies is the observation of the dependence of the
number of $0^+$ states as a function of valence nucleon numbers.
A particularly large number of low-lying states was interpreted as a signature of a
shape phase transition (Gd region), and the sharp drop of the number
of  low-lying $0^+$ states  was interpreted  as a result of proximity to the shell closure.
A particularly interesting result was obtained from the statistical analysis of
the distribution of $0^+$ energies: using the Brody distribution function suggests
that the spectrum of these excitations is intermediate between ordered and chaotic character.
More information from the (p,t) experiments, as well as on the  $0^+$ excitations
in even nuclei, was given in Refs.~\cite{Buc06,Lev09,Lev13}.
They report data on spins and cross sections for all states observed in the (p,t) reaction.
This allowed to extract information about the moments of inertia for the bands built
on the $0^+$ states.
These experimental studies contributed to the development of theoretical calculations,
which explain some of the features of the $0^+$ excitation spectra.
Some publications have dealt with the microscopic approach~\cite{LoI04,LoI05},
but the majority of studies used the phenomenological model of interacting
bosons (IBM)~\cite{Zam02,Sun03}.
These approaches have been used also in Ref.~\cite{Buc06,Lev09,Lev13}. Nevertheless, the nature
of multiple $0^+$ excitations in even nuclei is still far from being understood~\cite{She08}.

\begin{figure*}
   \epsfig{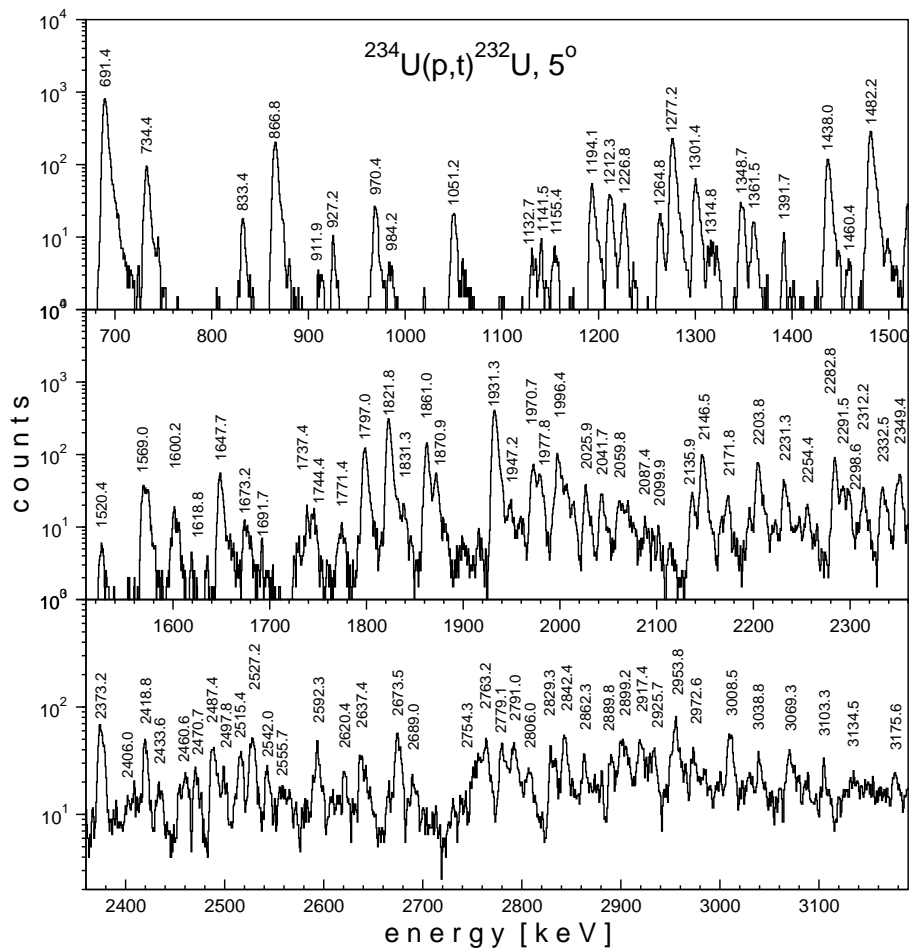}
    \caption{\label{fig:spec232u_5deg}
        Triton energy spectrum from the $^{234}$U(p,t)$^{232}$U reaction (E$_p$=25 MeV) in
        logarithmic scale for a detection angle of 5$^\circ$. Some strong lines
        are labelled with their corresponding level energies in keV.}
\end{figure*}

In this paper, we present the results of a careful and detailed analysis of the experimental
data from the high-resolution study of the $^{234}$U(p,t)$^{232}$U reaction. A short report
on this topic was presented in Ref.~\cite{Wir04}.
This analysis is similar to the one carried out for the nuclei $^{228}$Th and
$^{230}$Th~\cite{Lev09, Lev13}.
The nucleus $^{232}$U is located in the region of strong quadrupole deformation, where
stable reflection-asymmetric octupole deformations occur.
Information on excited states of $^{232}$U is rather scarce~\cite{Bro06}:
they have been studied via $^{232}$Pa $\beta^-$ decay, $^{232}$Np electron capture decay,
$^{236}$Pu $\alpha$ decay and via the  $^{230}$Th($\alpha$,2n$\gamma$) and
$^{232}$Th($\alpha$,4n$\gamma$) reactions.
The study of the (p,t) reaction adds to this information considerably: data are obtained
for 162 levels in the energy range up to 3.25 MeV.
Besides 0$^+$ states, where the number of reliable assignments could be increased from 9
to 13 states in comparison to the preliminary analysis in Ref.~\cite{Wir04},
information on the spins up to 6$^+$ for many other states was obtained.
Some levels are grouped into rotational bands, thus allowing to derive the moment of inertia
for some $0^+$, $2^+$ and $0^-$, $1^-$, $2^-$, $3^-$ bands.

\section{\label{sec:ExAnRe} Experiment details and   results}

\subsection{ \label{sec:det_exp} Details of the experiment}

The (p,t) experiment has been performed at the Tandem accelerator of the Maier-Leibnitz-Laboratory
of the Ludwig-Maximilians-Universit\"at and Technische Universit\"at M\"unchen.
A radioactive target of  100 $\mu$g/cm$^2$ $^{234}$U with half-life T$_{1/2}$ = 2.45$\cdot10^5$
years, evaporated on a 22 $\mu$g/cm$^2$ thick carbon backing, was bombarded with 25 MeV protons
at an intensity of 1-2 $\mu$A on the target.  The isotopic purity of the target was about 99\,\%.
The reaction products have been analyzed with the Q3D magnetic spectrograph  and then
 detected in a focal plane detector. The focal plane detector is a multiwire
 proportional chamber with readout of a cathode foil structure for
 position determination and dE/E particle identification \cite{Zan91,Wir01}.
 The acceptance of the spectrograph was 11 msr, except for the most forward angle of 5$^\circ$
 with an acceptance of 6 msr. The resulting triton spectra have a resolution of
 4--7 keV (FWHM) and are background-free. The experimental runs were normalized to the integrated
 beam current measured in a Faraday cup behind the target.
 The angular distributions of the cross sections were obtained from the
 triton spectra at twelve different laboratory angles from 5$^\circ$ to 50$^\circ$ in two sets:
 the first one with higher accuracy  for energies  up to 2350 keV and the second one
 with  somewhat lower accuracy for energies from 2200 to 3250 keV.

 A triton energy spectrum measured at a detection angle of 5$^\circ$ is shown in
 Fig.~\ref{fig:spec232u_5deg}. At this angle, the 0$^+$ states have comparatively
 large cross sections. The analysis of the triton spectra was performed using
 the program GASPAN \cite{Rie91}.  For the calibration of the energy scale,
the triton spectra from the reactions $^{184}$W(p,t)$^{182}$W and
 $^{186}$W(p,t)$^{184}$W  were measured at the same magnetic settings.
 The known levels in $^{232}$U \cite{Bro06} and the levels in $^{228}$Th known from
 the study \cite{Lev13} were also included in the calibration.

 The peaks in the energy spectra for all  twelve angles were identified for 162 levels.
 The information obtained for these levels is summarized in  Table~\ref{tab:expEI}.
 The energies and spins of the levels as derived from this study are compared
 to known energies and spins from \cite{Bro06}. They are given in the first four columns.
 The column labelled $\sigma_{\mbox{integ.}}$ gives the cross section integrated
 in the region from 5$^{\circ}$ to 50$^{\circ}$.  The column entitled
{$\sigma_{\mbox{exp.}}/\sigma_{\mbox{calc.}}$} gives the ratio of the integrated cross sections, obtained
from experimental values, from calculations in the DWBA approximation (see Sec. \ref{sec:DWBA}).
The last column lists the notations of the schemes used in the DWBA calculations: sw.jj means
one-step direct transfer of the $(j)^2$ neutrons in the (p,t) reaction; notations of the
multi-step transfers used in the DWBA calculations are displayed in Fig.~\ref{fig:schemes}.

\newcolumntype{d}{D{.}{.}{3}}
\renewcommand{\thefootnote}{\fnsymbol{footnote}}
\begin{longtable*}{ll l ccr  c c c}
\caption{\label{tab:expEI} \normalsize Energies of  levels in $^{232}$U, the level spin
assignments from the CHUCK3 analysis, the (p,t) cross sections integrated over the measured
values (i.e. 5$^\circ$ to 50$^\circ$) and the reference to the schemes used in the DWBA calculations
(see text for more detailed explanations).}\\
\hline\hline
\smallskip\\
\multicolumn{3}{c}{Level energy [keV]}&\multicolumn{2}{c}{$I^\pi$}&\enspace & \enspace {Ratio}&Way of
\smallskip\\
\multicolumn{2}{l}{This work}&\enspace\enspace NDS[17]&\enspace This work & NDS[17]& \enspace \enspace {$\sigma_{\mbox{integ.}}$}[{$\mu$b}]&
  \enspace\enspace {$\sigma_{\mbox{expt.}}/\sigma_{\mbox{calc.}}$}&fitting\\
\smallskip\\
\hline
\endfirsthead
\caption{Continuation}\label{tab:expEI}\\
\hline\hline
\smallskip\\
\multicolumn{3}{c}{Level energy [keV]}&\multicolumn{2}{c}{$I^\pi$}&\enspace & \enspace {Ratio}&Way of
\smallskip\\
\multicolumn{2}{l}{This work}&\enspace\enspace NDS[17]&\enspace This work & NDS[17]& \enspace \enspace {$\sigma_{\mbox{integ.}}$}[{$\mu$b}]&
  \enspace\enspace {$\sigma_{\mbox{expt.}}/\sigma_{\mbox{calc.}}$}&fitting\\

\smallskip\\
\hline
\endhead
\hline
\endfoot
\endlastfoot
~~~0.0 \it1  &&    ~~~~0.00    & $0^+$         & $0^+$            &183.90 \it58   &8.95 &sw.gg \\
~~47.6 \it1  &&    ~~47.573(8) & $2^+$         & $2^+$            &43.40 \it35    &50.5 &m1a.gg\\
~156.6 \it1  &&    ~156.566(9) & $4^+$         & $4^+$            &8.11  \it35    &1.20 &m1a.gi\\
~322.6 \it2  &&    ~322.69(7)  & $6^+$         & $6^+$            &5.55  \it40    &178&m2d.gg \\
~541.1 \it4  &&    ~541.1(1)   & $(8^+)$       & $8^+$            &0.42  \it20    &0.75&m2c.gg \\
~563.2 \it4  &&    ~563.194(7) & $1^-$         & $1^-$            &0.90  \it25    &0.12&m1a.gg \\
~628.8 \it4  &&    ~628.965(8) & $3^-$         & $3^-$            &2.70  \it33    &0.24&m3a.gg \\
~691.4 \it2  &&    ~691.42(9)  & $0^+$         & $0^+$            &35.00 \it70    &194 & sw.ii\\
~734.4 \it2  &&    ~734.57(5)  & $2^+$         & $2^+$            &21.68 \it65    &2.85&m1a.gi\\
~746.5 \it5  &&    ~746.8(1)   & $ $           & $(5^-)$          &0.35  \it18    &    &      \\
~833.4 \it2  &&    ~833.07(20) & $4^+$         & $4^+$            &3.21  \it23    &0.55&m1a.gg\\
~866.8 \it2  &&    ~866.790(8) & $2^+$         & $2^+$            &64.05 \it90    &8.15&m1a.gi\\
~911.9 \it4  &&    ~911.49(4)  & $3^+$         & $(3^+)$          &1.08  \it15    &6.75&m2a.gg\\
~914.5 \it9  &&    ~915.2(4)   & $ $           & $7^-$            &0.10  \it06    &    &      \\
~927.2 \it4  &&    ~927.3(1)   & $0^++2^+$     & $(0^+)$          &0.35  \it10    &1.45&sw.ii\\
~967.1 \it9  &&    ~967.6(1)   &               & $(2^+)$          &0.65  \it35    &    &     \\
~970.4 \it3  &&    ~970.71(7)  & $4^+$         & $(4^+)$          &4.65  \it32    &69.0&m1a.ij\\
~984.2 \it9  &&    ~984.9(2)   & $6^+$         & $6^+$            &0.40  \it12    &13.0&m2d.gg\\
1015.9 \it9  &&    1016.850(8) & $(2^-)$       & $2^-$            &0.12  \it07    &0.20 &m2f.gg \\
1051.2 \it3  &&    1050.90(1)  & $3^-$        & $3^-$            &3.45  \it25    &0.24 &m1a.gg\\
1060.8 \it8  &&                & $(3^-)$      & $ $              &0.32  \it12    &0.14 &m2a.gg\\
1097.6 \it8  &&    1098.2(4)   & $(4^-)$      & $(4^-)$          &0.10  \it06    &3.15 &m2a.gg\\
1132.7 \it3  &&    1132.97(10) & $2^+$        & $(2^+)$          &1.52  \it18    &0.15 &sw.gg \\
1141.5 \it4  &&                & $(1^-)$      & $ $              &1.02  \it15    &0.13 &m1a.gg\\
1155.4 \it4  &&                & $5^-$        & $ $              &1.38  \it35    &0.78 &m2e.gg\\
             &&                &or $3^+$      &                  &               &6.40 &m2a.gg\\
1173.0 \it6  &&    1173.06(17) & $2^-$        & $(2)^-$          &0.52  \it12    &6.12&m2a.gg \\
1194.1 \it3  &&    1194.0(2)   & $4^+$        & $(3+,4^+)$       &3.18  \it53    &1.65&m2a.gg \\
1212.3 \it3  &&    1211.3(3)   & $3^-$        & $3^-$            &6.60  \it55    &0.46&m1a.gg \\
1226.8 \it4  &&               & $4^+$        & $ $              &2.64  \it26    &0.33&m1a.gj \\
1264.8 \it3  &&               & $3^-$        & $ $              &2.42  \it22    &1.00&m2a.gg \\
1277.2 \it3  &&               & $0^+$        & $ $              &16.12 \it60    &0.45 &sw.gg \\
1301.4 \it3  &&               & $2^+$        & $ $              &3.00  \it25    &3.95&m1a.gg \\
1314.8 \it4  &&               & $6^+$        & $ $              &3.57  \it28    &13.3&m2e.gg \\
1321.8 \it5  &&               & $2^+$        & $ $              &0.57  \it20    &2.90&sw.jj  \\
             &&              &or  $3^-$      & $ $              &               &0.12&m3a.gg \\
1348.7 \it3  &&               & $(2^+)$      & $ $              &3.25  \it25    &14.5&sw.jj  \\
1361.5 \it4  &&               & $4^+$        & $ $              &1.00  \it16    &0.45&m2a.gg \\
             &&              &or  $3^-$      & $ $              &               &0.16&m3a.gg \\
1372.0 \it6  &&               & $2^+$        & $ $              &0.30  \it12    &0.03&m1a.ig\\
             &&               &or $6^+$      & $ $              &               &0.33&m2d.gg\\
1391.7 \it4  &&               & $4^+$        & $ $              &0.85  \it15    &0.12&m1a.gg \\
             &&              &or $5^-$       & $ $              &               &12.0&sw.ji \\
1438.0 \it3  &&               & $4^+$        & $ $              &12.72 \it55    &205.&m1a.ij \\
1460.4 \it6  &&               & $6^+$        & $ $              &0.85  \it15    &0.27&m2d.gg \\
1482.2 \it3  &&               & $0^+$        & $ $              &14.15 \it55    &27.25 &sw.ig\\
1489.2 \it4  &&               & $2^+$        & $ $              &4.18  \it50    &0.45 &sw.gg \\
1501.4 \it7  &&              & $3^-$        & $ $              &0.68  \it15    &41.8 &sw.jj \\
1520.4 \it4  &&              & $2^+$        & $ $              &6.85  \it33    &150.&m1a.ii \\
1552.8 \it8  &&              & $(3^+)$      & $ $              &0.83  \it15    &4.60&m2a.gg \\
1569.0 \it4  &&              & $0^+$        & $ $              &3.72  \it36    &8.15 &sw.ig \\
1572.9 \it6  &&              & $4^+$        & $ $              &3.45  \it36    &31.5 &sw.jj \\
1600.2 \it6  &&              & $2^+$        & $ $              &1.40  \it25    &1.50 &m1a.gg\\
1605.4 \it8  &&              & $4^+$        & $ $              &0.72  \it22    &6.15 &m1a.ij\\
1618.8 \it7  &&              & $2^+$        & $ $              &0.62  \it15    &0.08&m1a.ig \\
1633.8 \it6  &&              & $3^+$        & $ $              &1.35  \it25    &5.05&m2a.gg \\
             &&             &or  $6^+$        &                  &               &16.0&sw.jj  \\
1647.7 \it5  &&              & $2^+$        & $ $              &24.88 \it50    &3.15&m1a.gg \\
1673.2 \it5  &&              & $4^+$        & $ $              &1.72  \it25    &0.35&sw.gg  \\
1679.8 \it6  &&              & $1^-$        & $ $              &1.25  \it22    &0.04&m1a.gg \\
             &&             &or  $3^-$        & $ $              &               &0.06&m3a.gg \\
1691.7 \it6  &&              & $(6^+)$      & $ $              &1.05  \it18    &4.20&m2e.gg \\
1700.5 \it8  &&              & $6^+$        & $ $              &0.85  \it18    &0.20&sw.gg \\
1728.5 \it6  &&              & $(4^+)$      & $ $              &1.08  \it22    &0.20&m1a.gg \\
1737.4 \it5  &&              & $(6^+)$      & $ $              &3.25  \it33    &0.35&m3a.gg \\
1744.4 \it5  &&              & $4^+$        & $ $              &4.86  \it40    &0.85&m1a.gg \\
             &&              &or  $5^-$     &                  &               &2.70 &m2e.gg \\
1758.9 \it9  &&              & $(5^-)$      & $ $              &0.69  \it15    &19.0&sw.ii  \\
1771.4 \it8  &&              & $ $          & $ $              &2.96  \it26    &    &       \\
1790.8 \it7  &&              & $6^+$        & $ $              &2.05  \it28    &7.50&m2e.gg \\
1797.0 \it5  &&              & $0^+$        & $ $              &10.15 \it65    & 11.0&sw.ii \\
1802.5 \it9  &&              & $(4^+)$      & $ $              &0.88  \it45    &9.80 &sw.ij\\
1821.8 \it5  &&              & $0^+$        & $ $              &20.65 \it70    &27.2&sw.ii  \\
1831.7 \it5  &&              & $(2^+)$      & $ $              &1.38  \it30    & 0.22&m1a.gg\\
1838.6 \it6  &&              & $2^+$        & $ $              &1.16  \it26    &1.40 &m1a.gg\\
1861.0 \it5  &&              & $0^+$        & $ $              &9.63 \it50     &11.2 &sw.ig \\
1870.9 \it5  &&              & $2^+$        & $ $              &14.18 \it65    &1.70&m1a.gi \\
1880.8 \it5  &&              & $6^+$        & $ $              &2.08  \it35    &0.18&m3a.gg \\
1900.0 \it6  &&              & $ $          & $ $              &1.25  \it25    & & \\
1915.2 \it8  &&              & $6^+$        & $ $              &1.85  \it30    &7.40 &m2d.gg\\
1931.3 \it5  &&              & $0^+$        & $ $              &29.55 \it75    &140 &sw.ig  \\
1947.2 \it8  &&              & ($0^+$)      & $ $              &1.52  \it30    &7.60&sw.ig  \\
1957.4 \it8  &&              & $6^+$        & $ $              &3.05  \it35    &11.9 &m2d.gg\\
1970.7 \it5  &&              & $2^+$        & $ $              &25.65 \it95    &2.70 &sw.ig \\
1977.8 \it5  &&              & $2^+$        & $ $              &20.05 \it90    &2.25 &m1a.gg\\
1996.4 \it5  &&              & $4^+$        & $ $              &15.20 \it65    &108 &sw.ij  \\
2004.9 \it6  &&              & $4^+$        & $ $              &4.70  \it50    &46.0&sw.ij  \\
2011.6 \it6  &&              & $ $          & $ $              &2.30  \it65    &    &       \\
2025.9 \it6  &&              & $0^+$      & $ $              &2.96  \it35    &19.0&sw.ii  \\
2041.7 \it5  &&              & $2^+$        & $ $              &10.85 \it55    &1.20&m1a.gg \\
2059.8 \it5  &&              & $2^+$        & $ $              &10.65 \it55    &1.20&m1a.gg \\
2068.6 \it5  &&              & $4^+$        & $ $              &3.40  \it45    &0.42& sw.gg \\
2073.4 \it9  &&              &              &                  &1.30  \it40    &    &       \\
2087.4 \it6  &&              & $5^-$        & $ $              &               &1.95&m2e.gg \\
             &&             &or  $6^+$      & $ $              &               &53.0&sw.jj \\
2094.2 \it8  &&              &              &                  &1.20  \it40    &    &       \\
2099.9 \it6  &&              & $6^+$        & $ $              &1.62  \it35    &21.3&sw.jj  \\
2135.9 \it5  &&              & $4^+$        & $ $              &4.22  \it55    &32.4&m1a.ij \\
2146.5 \it5  &&              & $2^+$        & $ $              &39.80 \it98    &4.60&m1a.gg \\
2171.8 \it5  &&              & $2^+$        & $ $              &8.81  \it55    &1.00&sw.gg  \\
2194.6 \it5  &&              & $2^+$        & $ $              &4.08  \it55    &0.45&sw.gg  \\
2203.8 \it5  &&              & $2^+$        & $ $              &29.20 \it95    &3.15&sw.gg  \\
2221.3 \it9  &&              &              &                  &1.22  \it45    &    &       \\
2231.3 \it5  &&              & $4^+$        & $ $              &10.38 \it58    &67.0&m1a.ij \\
2235.9 \it5  &&              & $ $          & $ $              &2.00  \it60    &    &       \\
2246.2 \it5  &&              & $ $          & $ $              &1.55  \it55    &    &       \\
2254.4 \it5  &&              & $6^+$        & $ $              &5.35  \it45    &19.2& m2d.gg\\
2282.8 \it5  &&              & $2^+$        & $ $              &29.30 \it70    &3.30&m1a.gg \\
2291.5 \it5  &&              & $2^+$        & $ $              &11.23 \it90    &1.28&m1a.gg \\
2298.6 \it5  &&              & $2^+$        & $ $              &4.73  \it85    &1.35&sw.gg  \\
2312.2 \it6  &&              & $4^+$        & $ $              &3.95  \it75    &28.5&m1a.ij \\
2332.5 \it6  &&              & $2^+$        & $ $              &8.98  \it65    &1.10&m1a.gg \\
2349.4 \it6  &&              & $2^+$        & $ $              &17.27 \it86    &2.05&m1a.gg \\
2373.2 \it6  &&              & $2^+$        & $ $              &19.15 \it90    &2.25&m1a.gg \\
2397.7 \it6  &&              & $2^+$        & $ $              &2.48  \it48    &0.21&sw.gg  \\
2406.0 \it6  &&              & $6^+$        & $ $              &2.80  \it68    &12.8&m2e.gg \\
             &&              &or $5^-$        &                  &               &110 &sw.ij  \\
2412.4 \it6  &&              & $2^+$        & $ $              &3.67  \it90    &0.35&m1a.gg \\
2418.8 \it5  &&              & $2^+$        & $ $              &11.88 \it92    &1.45&m1a.gg \\
2433.6 \it5  &&              & $3^-$        & $ $              &2.61  \it47    &3.80&m2a.gg  \\
2454.2 \it5  &&              & $(3^-)$      & $ $              &1.67  \it96    &2.60&m2a.gg \\
2460.6 \it5  &&              & $3^-$        & $ $              &4.14  \it98    &5.20&m2a.gg  \\
             &&             &or  $6^+$        & $ $              &               &0.45&m3a.gg \\
2470.7 \it6  &&              & $(3^-)$      & $ $              &3.32  \it63    &4.90&m2a.gg  \\
2487.4 \it5  &&              & $3^-$        & $ $              &5.96  \it68    &8.90&sw.gg \\
2497.8 \it6  &&              & $(4^+)$      & $ $              &2.82  \it58    &21.4&m1a.ij \\
2515.4 \it6  &&              & $(3^-)$      & $ $              &8.21  \it74    &10.8&m2a.gg \\
2527.2 \it6  &&              & $4^+$        & $ $              &7.69  \it73    &46.0&m1a.ij \\
2542.0 \it7  &&              & $2^+$        & $ $              &5.88  \it69    &0.67&m1a.gg \\
2555.7 \it8  &&              & $(4^+)$      & $ $              &1.98  \it62    &11.2&m1a.ij \\
2564.7 \it8  &&              & $(3^-)$      & $ $              &1.82  \it62    &2.60&m2a.gg \\
2582.9 \it8  &&              & $ $          & $ $              &1.85  \it66    &    &       \\
2592.3 \it7  &&              & $4^+$        & $ $              &5.78  \it71    &32.9&m1a.ij \\
2598.7 \it9  &&              &              &                  &1.20  \it40    &    &       \\
2608.0 \it9  &&              & $2^+$        & $ $              &2.15  \it53    &0.32&m1a.gg \\
2620.4 \it6  &&              & $ $          & $ $              &2.75  \it62    &    &       \\
2637.4 \it6  &&              & $6^+$        & $ $              &6.84  \it87    &0.65&m3a.gg \\
             &&              &or $5^-$      &                  &               &180 &sw.ij  \\
2642.0 \it7  &&              & $ $          & $ $              &1.78  \it80    &    &       \\
2664.6 \it7  &&              & $4^+$        & $ $              &1.75  \it45    &11.5&m1a.ij \\
2673.5 \it7  &&              & $2^+$        & $ $              &9.11  \it76    &1.08&m1a.gg \\
2689.0 \it8  &&              & $2^+$        & $ $              &3.45  \it58    &0.38&m1a.gg \\
2754.3 \it7  &&              & $4^+$        & $ $              &4.89  \it68    &29.5&m1a.ij \\
2763.2 \it6  &&              & $(3^-)$      & $ $              &6.08  \it70    &8.95&sw.gg  \\
2779.1 \it6  &&              & $2^+$        & $ $              &7.21  \it70    &0.92&m1a.gg \\
2791.0 \it7  &&              & $2^+$        & $ $              &8.90  \it75    &1.18&m1a.gg \\
2806.0 \it7  &&              & $2^+$        & $ $              &4.50  \it62    &0.66&m1a.gg \\
2829.3 \it7  &&              & $4^+$        & $ $              &5.98  \it65    &34.5&m1a.ij \\
2842.4 \it7  &&              & $4^+$        & $ $              &7.12  \it75    &38.0&m1a.ij \\
2850.6 \it7  &&              & $6^+$        & $ $              &2.63  \it63    &11.9&m2d.gg \\
2862.3 \it7  &&              & $3^-$        & $ $              &4.58  \it63    &6.75&sw.gg  \\
2878.3 \it7  &&              & $(6^+)$      & $ $              &4.45  \it63    &345 &sw.ig  \\
2889.8 \it6  &&              & $4^+$        & $ $              &4.60  \it72    &26.0&m1a.ij \\
2899.2 \it7  &&              & $(4^+)$      & $ $              &5.1   \it15    & 0.75& sw.gg\\
2905.8 \it7  &&              & $ $          & $ $              &3.5   \it16    &    &       \\
2917.4 \it7  &&              & $0^+$        & $ $              &5.58  \it92    &8.95& sw.ji \\
2925.7 \it8  &&              & $(6^+)$      & $ $              &3.4   \it18    &14.5&m2d.gg \\
2931.5 \it7  &&              & $(5^-)$      & $ $              &5.8   \it20    &61.0&sw.ij  \\
2953.5 \it8  &&              & $4^+$        & $ $              &9.60  \it80    &58.5&m1a.ij \\
2959.7 \it7  &&              & $(2^+)$      & $ $              &3.05  \it60    &8.90&sw.ig  \\
2972.6 \it8  &&              & $2^+$        & $ $              &4.70  \it65    &0.48&m1a.gg \\
2984.2 \it8  &&              & $ $          & $ $              &1.09  \it50    &    &       \\
2998.7 \it8  &&              & $(2^+)$      & $ $              &2.40  \it54    &7.40&sw.ig  \\
3008.5 \it8  &&              & $(3^-)$      & $ $              &2.80  \it65    &0.52&m3a.gg \\
3028.8 \it8  &&              & $4^+$        & $ $              &2.05  \it55    &12.0&m1a.ij \\
3038.8 \it8  &&              & $(5^-)$      & $ $              &4.40  \it63    &145 &sw.ij  \\
3058.3 \it9  &&              & $(6^+)$      & $ $              &1.00  \it50    &91.0&sw.ig  \\
3069.3 \it8  &&              & $3^-$        & $ $              &5.04  \it75    &6.75&m2a.gg \\
3075.7 \it9  &&              & $(5^-)$      & $ $              &1.88  \it65    &61.5&sw.ij  \\
3087.5 \it9  &&              & $2^+$        & $ $              &1.70  \it65    &0.20&m1a.gg \\
3103.3 \it9  &&              & $(4^+)$      & $ $              &1.95  \it55    &10.2&m1a.ij \\
3134.5 \it9  &&              & $(4^+)$      & $ $              &1.68  \it55    &9.90&m1a.ij \\
3149.1 \it9  &&              & $2^+$        & $ $              &1.85  \it55    &41.8&sw.ij  \\
             &&             &or  $3^-$      & $ $              &               &2.60&m2a.gg \\
3175.6 \it8  &&              & $(2^+)$      & $ $              &1.96  \it55    &42.0&sw.ij  \\
\hline \\
\end{longtable*}

\subsection{\label{sec:DWBA} DWBA analysis}

We assume that a {\it(lj)} pair transferred in the (p,t) reaction  is  coupled to spin zero,
and that  the overall shape of the angular
distribution of the cross section is rather independent of the specific structure of the
individual states, since the wave function of the outgoing tritons is restricted to the nuclear
exterior and therefore to the tails of the triton form factors.  At the same time, cross sections
for different orbits have to differ strongly in magnitude. To verify this assumption, DWBA
calculations of angular distributions for different $(j)^2$ transfer configurations to states
with different spins were carried out in our previous paper \cite{Lev09}. Indeed, the magnitude
of the cross sections differs  strongly for different orbits, but  the shapes of
calculated angular distributions are very similar. Nevertheless, they depend
 to some degree on the transfer configuration, the most pronounced being found for
 the 0$^+$ states, which is confirmed  by the experimental angular distributions.
 This is true for most of the {\it(lj)} pairs and only for the case of a one-step transfer.
No complication of the angular distributions is expected for the excitation
of  $0^+$ states, which  proceeds predominantly via a one-step process. This is not the case for the
excitation of states with other spins, where the angular distribution may be altered due to
inelastic scattering (coupled channel effect), treated here as multi-step processes.
Taking into account these circumstances allows for a reliable  assignment of  spins for most
of the excited states in the final nucleus $^{232}$U by  fitting the  angular distributions
obtained in the DWBA calculations to the experimental ones. The assignment of a single spin has not been possible
 only in a few cases,  for which two or even three spin values are allowed.

\begin{figure}
\begin{center}
\epsfig{file=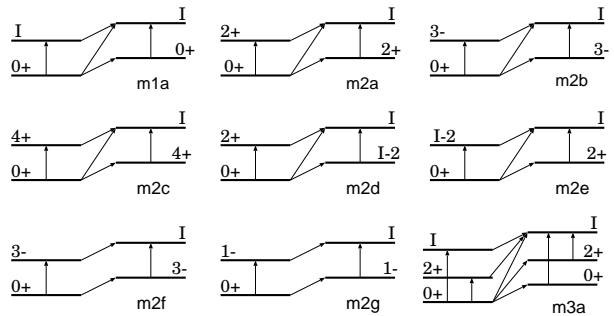, width=8 cm,angle=0}
    \caption{\label{fig:schemes}
    Schemes of the CHUCK3 multi-step
    calculations tested with spin assignments of  excited states
    in $^{232}$U (see Table~\ref{tab:expEI}).}
\end{center}
\end{figure}


 The magnitude and shape of the DWBA cross section angular distributions depends on
 the chosen potential parameters. We used the optical potential parameters suggested
 by Becchetti and Greenlees \cite{Bec69}  for protons and by Flynn {\it et al.} \cite{Fly69}
 for tritons. These parameters have been tested via their description of angular distributions
 for the ground states of $^{228}$Th, $^{230}$Th and $^{232}$U  \cite{Wir04}.
 Minor changes of the parameters for tritons were needed only for some $3^-$ states,
 particularly for the states at 628.8, 1051.2 and 1212.3 keV. For these states, the triton
potential parameters suggested by Becchetti and Greenlees \cite{Bec71} have been used.
 For each state the binding energies of the two neutrons are calculated to match the outgoing
 triton energies. The corrections to the reaction energy are introduced depending
 on the excitation energy. For more details see \cite{Lev09}.

\begin{figure}
\begin{center}
\epsfig{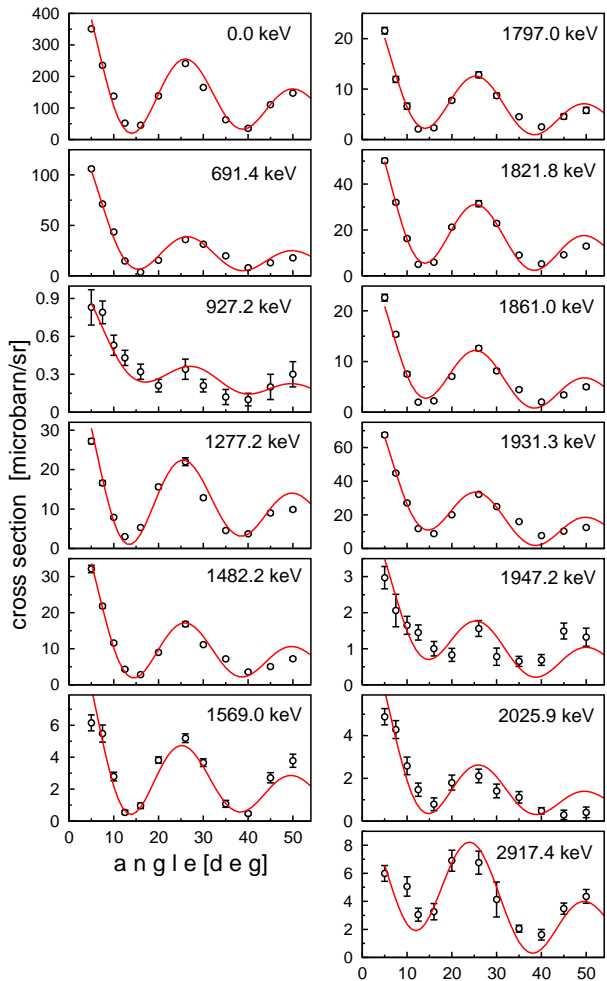}
    \caption{\label{fig:angl_distr_0+} (Color online)
    Angular distributions of assigned $0^+$ states in $^{232}$U and
     their fit with  CHUCK3 one-step calculations. The  transfer
     configurations used in the calculations for the best fit are
    given in Table~\ref{tab:expEI}. See text for further information. }
\end{center}
\end{figure}

 The coupled-channel approximation (CHUCK3 code of Kunz \cite{Kun}) was used in previous \cite{Lev09,Lev13}
 and present calculations. The best reproduction of the angular distribution for the ground
 state and for the 1277.2 keV state was obtained for the transfer of the $(2g_{9/2})^2$ configuration
 in the one-step process. This orbital is close to the  Fermi surface and was considered in previous
 studies \cite{Lev09,Lev13} as the most probable one in the transfer process.  But for $^{232}$U,
 a better reproduction of the angular distributions for other $0^+$ states is obtained  for the configuration
$(1i_{11/2})^2$,  also near the Fermi surface, alone or in combination with the $(2g_{9/2})^2$ configuration.
The only exception is the state at 2917.4 keV, for which the experimental angular distribution can be
fitted only by the calculated one for the transfer of the  $(1j_{15/2})^2$ neutron configuration.

\begin{figure*}
\begin{center}
\epsfig{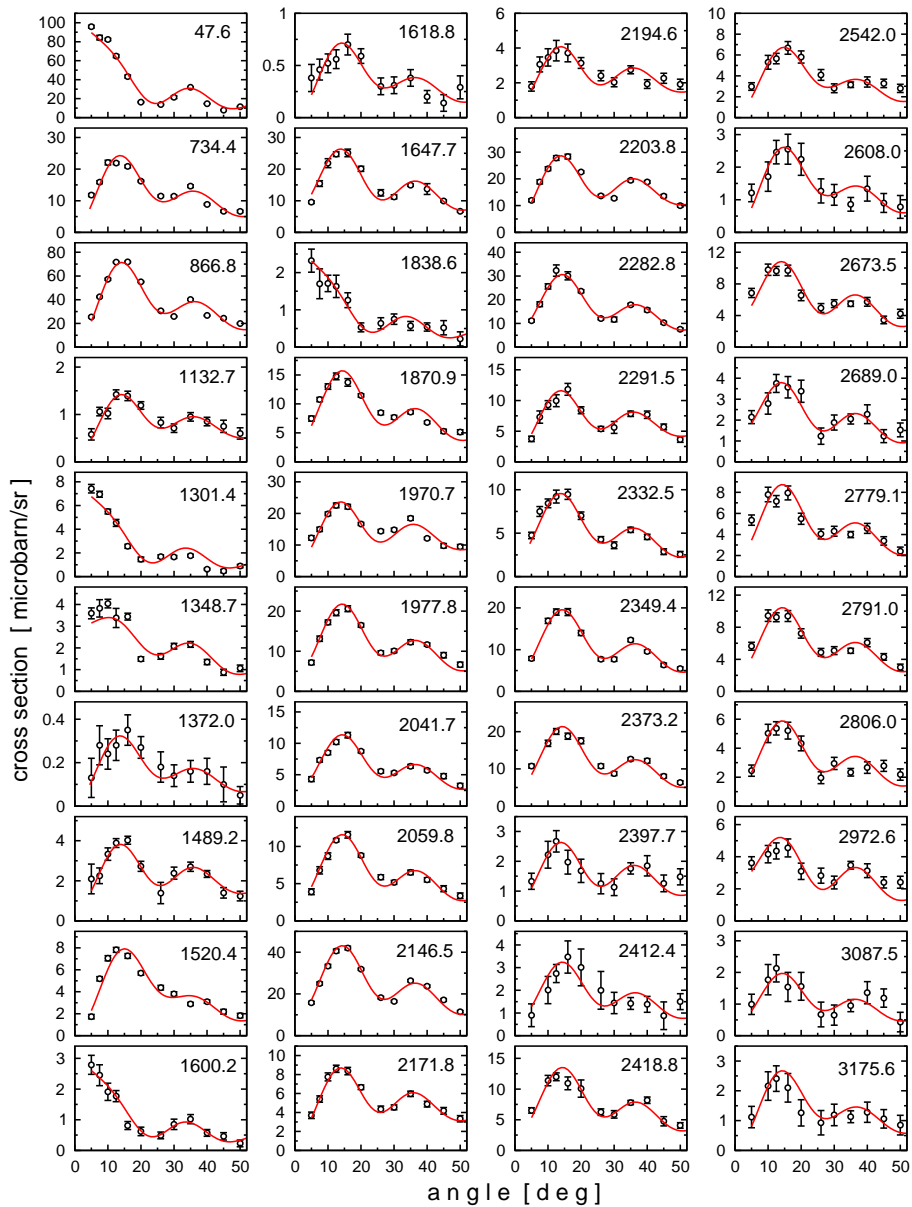}
    \caption{\label{fig:angl_distr_2}(Color online) Angular distributions of assigned
    $2^+$ states in $^{232}$U and their fit with  CHUCK3  calculations.
     The  $(ij)$ transfer configurations and schemes used
    in the calculations for the best fit are given in Table~\ref{tab:expEI}.}
\end{center}
\end{figure*}

Results of  fitting the angular distributions for the states assigned as $0^+$ excitations are
shown in Fig.~\ref{fig:angl_distr_0+}.  The agreement between the fit and the data is
excellent for most of the levels. Remarks are needed only for the level at  927.2 keV.
The existence of this state and  the state at the energy of 967.7 keV  was established
by the $\gamma$ energies and the coincident  results at the $\alpha$ decay of $^{236}$Pu \cite{Ard94}.
Strong evidence has been obtained that these states have
 spins $0^+$ and $2^+$ and are the members of a $K^{\pi} = 0^+$ band. At the same time, it was noted that
 the occurrence of a 927.7 keV $\gamma$ ray is in contradiction with the $0^+$ assignment for this state
 if this $\gamma$ ray corresponds to a ground-state transition from the 927.7 keV state.
 Alternatively, this transition should be placed in another location.
 The measured (p,t) angular distribution for the 927.7 keV state strongly peaks
 in the forward direction, which is typical for the L = 0  transfer but the lack of a deep minimum at
 about 14 degrees  contradicts the $0^+$ assignment. The assumption that a doublet with
 spins $0^+$ and $2^+$  occurs at the energy of 927.7 keV seems to be a unique explanation of
 the experimental data. The angular distribution in
 this case  is fitted by the calculated one satisfactorily as one can see in Fig.~\ref{fig:angl_distr_0+}.
 In order to obtain a satisfactory fit one has to assume a population of the 2$^+$ state at about 1/3 of
 the population of the 0$^+$ state. Thus we can make firm $0^+$ assignments for 12 states for energy excitations below 3.25 MeV,
 in comparison with 9 states found in the preliminary analysis of the
 experimental data \cite{Wir04}. The assignment for the 1947.2 keV level is tentative.
 We can compare 24 $0^+$ states in $^{230}$Th and 18 $0^+$ states in $^{228}$Th
 with only 13 $0^+$ states in $^{232}$U in the same energy region.

\begin{figure*}
\begin{center}
 \epsfig{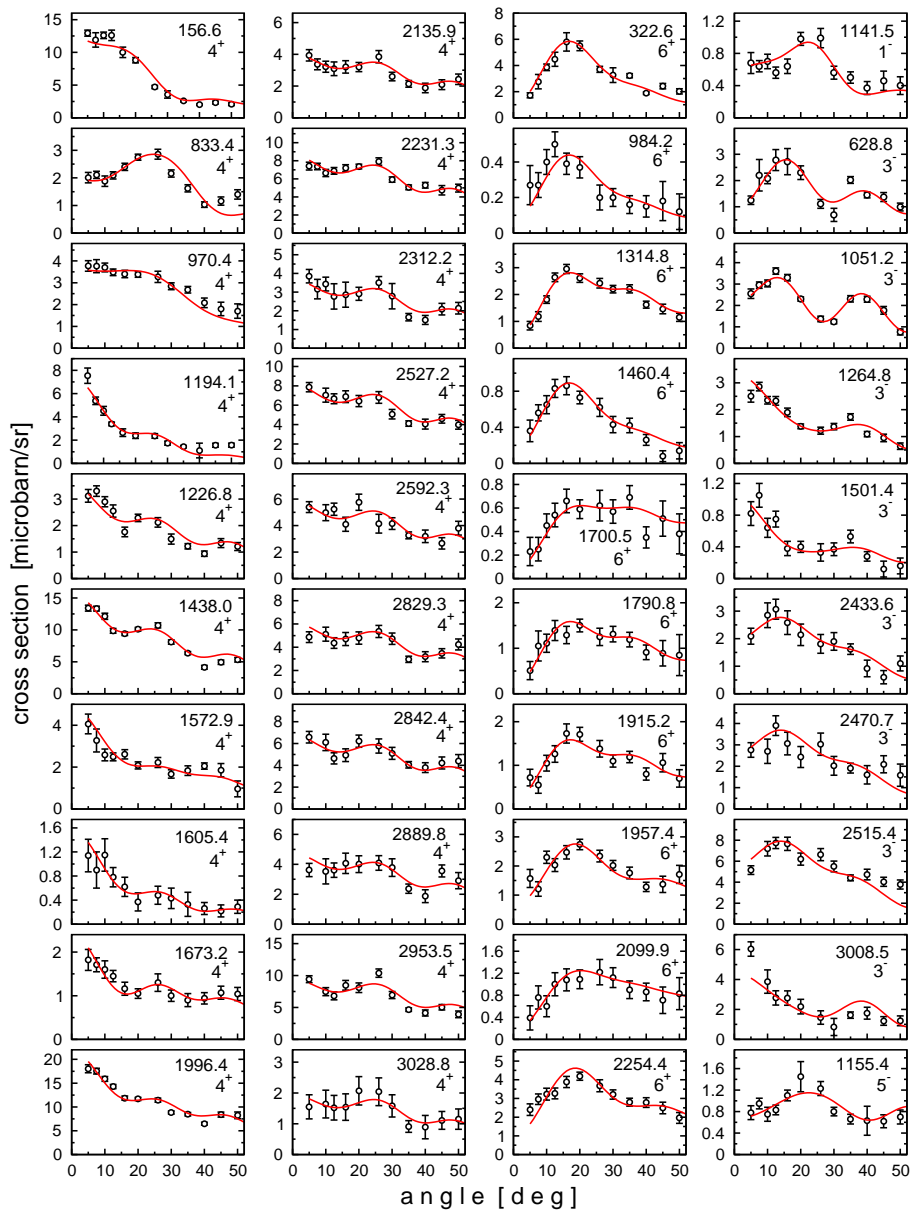}
    \caption{\label{fig:angl_distr_4635}(Color online) Angular distributions
    of some assigned states in $^{232}$U and their fit with
    CHUCK3 calculations: $4^+$ and $6^+$ with positive parity
    and $1^-$, $3^-$ and $5^-$ with negative parity.
    The  $(ij)$ transfer configurations and schemes used
    in the calculations for the best fit are given in Table~\ref{tab:expEI}.}
\end{center}
\end{figure*}

\begin{figure}
\begin{center}
 \epsfig{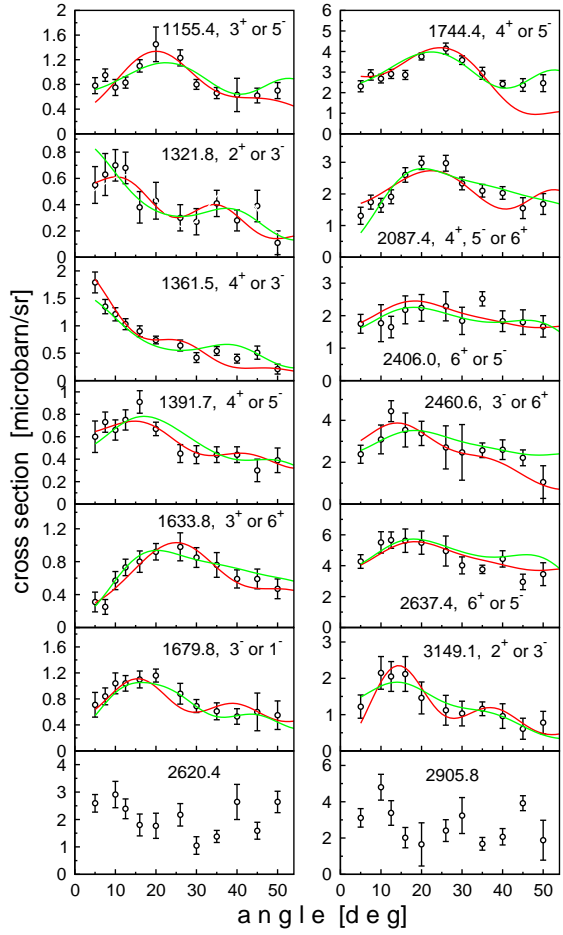}
    \caption{\label{fig:two_fit}(Color online) Angular distributions
    for some  states in $^{232}$U for which fitting of the calculated distributions
      for a unique spin is doubtful or not possible. The first spin is indicated
      by red color.}
\end{center}
\end{figure}

\begin{figure}
\begin{center}
\epsfig{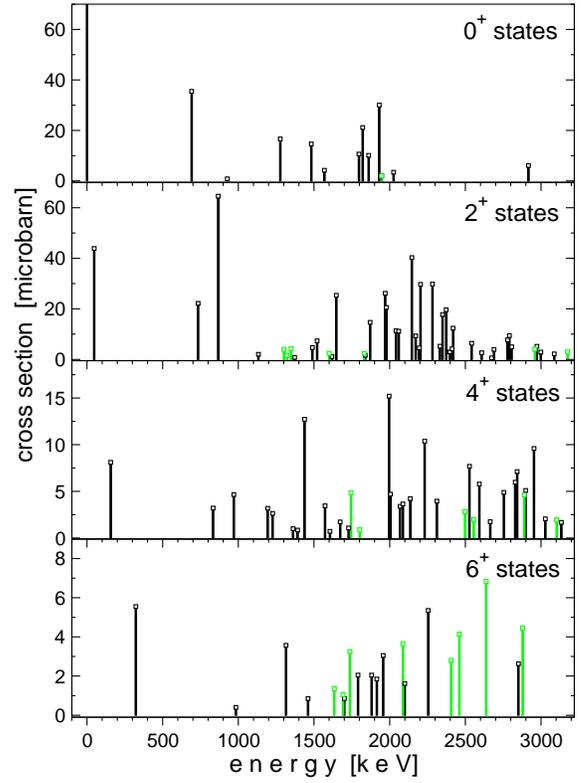}
    \caption{\label{fig:strength}(Color online) Experimental  distribution of
    the (p,t) strength integrated in the angle region 5$^\circ$  - 50$^\circ$
    for 0$^+$, 2$^+$, 4$^+$ and 6$^+$ states in $^{232}$U. Green lines represent
    tentative assignments.}
\end{center}
\end{figure}

The main goal of many studies using two-neutron transfer in the regions of rare earth, transitional
and spherical nuclei \cite{Les02,Mey06,Buc06,Bet09,Ili10,Ber13} was  to collect information only about the $0^+$ states, their energies
and excitation cross sections. At the same time the states with non-zero spin are intensively excited
in the (p,t) reaction and information about them can be obtained from the analysis of the angular distributions.
The main features of the angular distribution shapes for 2$^+$, 4$^+$ and 6$^+$ states are even more weakly
dependent on the transfer configurations  only in the case of one-step transfer.
Therefore the $(2g_{9/2})^2$, $(1i_{11/2})^2$ and $(1j_{15/2})^2$ configurations alone or in combination,
were used in the calculations for these states.
The one-step transfer calculations give a satisfactory fit of
angular distributions for about 30\,\% of the states with spins different from  $0^+$ and  the inclusion of multi-step excitations for about
70\,\% of the states is needed.  As in the Th isotopes
\cite{Lev09,Lev13}, multi-step excitations have to be included to fit the angular distributions
already for the $2^+$, $4^+$ and $6^+$ states of the g.s. band.
At least a small admixture of multi-step transfer for
most of the other states is required  to get a good agreement with experiment. Fig.~\ref{fig:schemes} shows
the schemes of the multi-step excitations, tested for every state in those cases,
where one-step transfer did not provide a successful fit.
Fig.~\ref{fig:angl_distr_2} demonstrates the quality of the fit of  different-shaped
angular distributions at the excitation of states with  spin $2^+$ by calculations assuming
one-step and one-step plus two-step excitations. Results of similar fits for the
states assigned as $4^+$, $6^+$ and $1^-$, $3^-$, $5^-$ excitations are shown in
Fig.~\ref{fig:angl_distr_4635}. At the same time, for a number of states, possibly due to a lack
of statistical accuracy, a good fit of the calculated angular distributions to the experimental
ones can not be achieved for a unique spin of the final state and therefore  uncertainties
remain in the spin assignment for such states. Some of them are demonstrated in Fig.~\ref{fig:two_fit}.

The spins and parities resulting from such fits are presented in Table~\ref{tab:expEI},
together with other experimental data.  Figure \ref{fig:strength} summarizes the (p,t)
strengths integrated over the angle region 5$^\circ$  - 50$^\circ$ for positive parity states.
The sixth column in  Table~\ref{tab:expEI} displays the ratio $\sigma_{exp}/\sigma_{cal}$.
Calculated cross sections for the specific transfer configurations
differ very strongly.  If  the microscopic structure of the
excited states is known, and thus the relative contribution of the specific $(j)^2$
transfer configurations to each of these states, these relationships are considered as
spectroscopic factors. A perfect fit of the experimental angular distributions may mean
that the assumed configurations in the calculations  correspond to the major
components of the real configurations. Therefore,  at least the order of magnitude for
the ratio $\sigma_{exp}/\sigma_{cal}$ corresponds
to the  actual spectroscopic factors with the exception of too large values, such as in the case of the
$(1i_{11/2})^2$ transfer configurations used in the calculation for some $0^+$ and even $2^+$
and $4^+$ states. Surprisingly, the shape just for this neutron configuration
gives the best agreement with  experiment for the mentioned states.

A few additional comments have to be added for the region where data about the spins and parities are known
from the analysis of $\gamma$ spectra \cite{Bro06}.
The angular distributions for some  states are  very different from those calculated for
the one-step transfer. Therefore, they were used as examples  for other  states at higher energies
in the analysis of the angular distributions.
As already  noted the difference is significant already for the 2$^+$ and 4$^+$ states of the g.s. band.
For example, the angular distribution for the 2$^+$ state at 47.6 keV can be used as a example for the states
at 1301.4, 1600.2 and 1838.6 keV.
From the two spins 3$^+$ and 4$^+$
proposed for the state  at 1194.1 keV in the analysis of the $\gamma$ spectra \cite{Bro06}, our data clearly
confirm  the spin 4$^+$. Then the angular distribution for this state can  serve as an example
for the states at 1361.5 keV and 1604.9 keV. Importantly,
the angular distributions for some 2$^+$ and 4$^+$ states have a feature typical for the excitation of 0$^+$ states,
namely a strong peak at small angles.

The angular distribution  for the 4$^+$ state at 833.4 keV, which is known from the $\gamma$ spectroscopy,
is very different from the one for the one-step transfer. It was used as an example for the assignment of
spins of the states at 1728.5 keV and 1744.4 keV with similar angular distributions.
Similarly, the angular distribution for the 1$^-$ state
at 563.2 keV can serve as an example for the state at 1141.3 keV.
The angular distribution for the state 3$^-$ at 628.8 keV  not only differs from the one calculated
for one-step transfer and can be described by the scheme m3a.gg, but it is very similar
to the angular distribution for the 2$^+$ state at one-step transfer. Therefore, for all  states
with similar experimental distributions,  the calculated angular distributions for the spin 2$^+$
 and 3$^-$ were tested during fitting procedure, using the scheme m3a.gg.

The states with unnatural parity populated via two neutron transfer, such as 3$^+$ at 911.9 keV
and 2$^-$ at 1173.0 keV, represent a special case. Assignments based on the $\gamma$-spectra analysis are tentative. As one can see
from Fig.~\ref{fig:two_fit}  these spins and parities are confirmed by fitting the angular distributions.
Spin 3$^+$ for the states at 1552.8 and 1633.8 keV is attributed taking into account also the similarity
of their angular distributions with those for the state at 911.9 keV.
The state at 1015.9 keV is excited weakly, but the angular distribution measured with small statistics
does not contradict the assignment of spin 2$^-$. The same is true for the state at 1097.6 keV with spin 4$^-$.

\section{\label{Disc} Discussion}

\subsection{\label{sec:Col_mom} Collective bands and moments of inertia in $^{232}$U}

\begin{figure}
\begin{center}
\epsfig{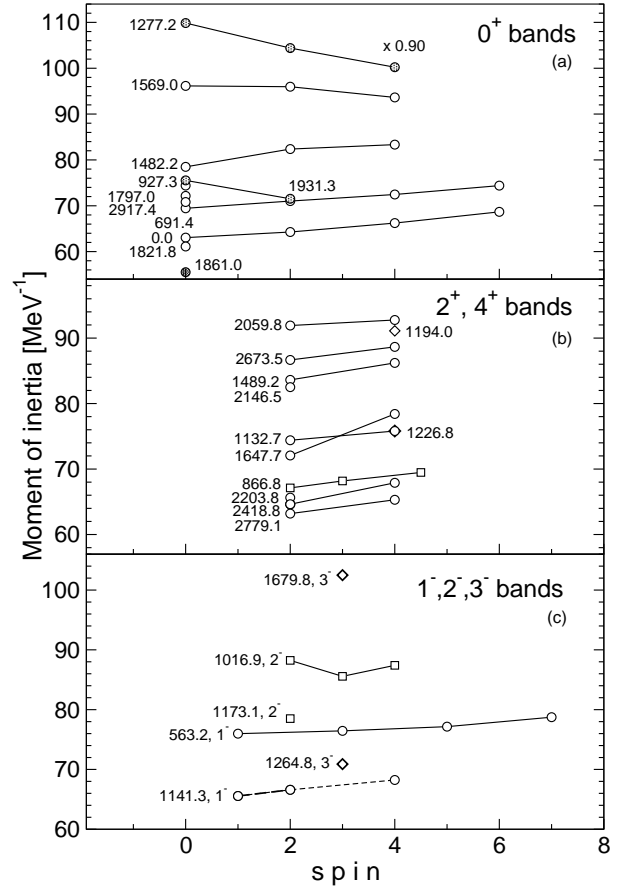} \caption{\label{fig:moments} Moments of inertia for
the bands in $^{232}$U, as assigned from the angular distributions
  from the $^{234}$U(p,t)$^{232}$U reaction.  Values of $J/\hbar^2$ are given.  In the graph,
  they are placed at spins of the first state from the two  used in the calculations of every value.}
\end{center}
\end{figure}

Aiming to get more information on the excited states in $^{232}$U,  especially
on the moments of inertia for the 0$^+$ states, we have attempted to identify
those sequences of  states,  which show the characteristics of a rotational band
structure. An identification of the states attributed to rotational bands
can be made on the basis of the following conditions:
a) the angular distribution for a  band member candidate  is fitted by  DWBA calculations
for its expected spin;
b) the transfer cross section in the (p,t) reaction  to states in the potential band has to
decrease with increasing  spin;
c) the energies of the states in the band can be fitted
approximately by the expression for a rotational band   $E = E_0 +
AI(I+1)$ with a small and smooth variation of the inertial
parameter $A$. Collective bands identified in such a way
 are listed in Table~\ref{tab:bands}. The procedure can be justified, since
 some sequences meeting the above
criteria are already known  to be rotational bands from gamma-ray spectroscopy
\cite{Bro06}. In Fig.~\ref{fig:moments}
we present moments of inertia (MoI) obtained by fitting the level
energies of the bands displayed in Table~\ref{tab:bands}  by the expression $E = E_0 + AI(I+1)$
for close-lying levels,  i.e. they were determined for  band members  using the ratio of $\Delta
E$ and $\Delta [I(I+1)]$, thus saving the spin dependence of the MoI.

\textsl{Negative parity states.} Unlike the thorium isotopes \cite{Lev09,Lev13}, some uncertainties in
 formation of the bands are met for $^{232}$U. At the beginning a few comments follow about
 the lowest negative-parity states, usually interpreted as of octupolar vibrational character.
 They are  one-phonon octupole excitations,  forming a quadruplet of states with
 $K^{\pi} = 0^-, 1^-, 2^-, 3^-$ and are the bandheads for the rotational bands.
 The $K^{\pi} = 0^-$ band is reliably established \cite{Bro06} and confirmed by the present study.
 There are two states with $J^{\pi} = 2^-$ at 1016.8 and 1173.1 keV, which may be members
 of bands with $K^{\pi} = 1^-$  and $K^{\pi} = 2^-$. The level  at 1146.3 keV  has
 been proposed as a bandhead of the $K^{\pi} = 1^-$ band from the observation of $\gamma$ ray
  with this energy \cite{Wei72}. The corresponding line in the triton spectra is absent.
 After our firm assignment of  spin 4$^+$ to the state at 1194.1 keV,  this proposal has
 to be rejected since the 1146.3 keV transition should  be referred to the decay of this
 level to the  2$^+$ level at 47.6 keV. At the same time a line in the triton spectra is observed
 at 1141.5 keV, the spin of corresponding state is assigned tentatively as 1$^-$.
Considering this level as the bandhead for the $1^-$ band with two levels  known from previous
studies \cite{Bro06}, the moments of inertia can be calculated.
The procedure described above can not be applied in this case because of the mixing by
the Coriolis interaction. A simplified expression for the band energies can be used in the analysis
(for details see \cite{Lev13})
\begin{eqnarray}
E(I,K^{\pi}=1^-) \sim E_1 +(A_1+B)I(I+1)\enspace \makebox{for $I$ odd}  \nonumber \\
E(I,K^{\pi}=1^-) \sim E_1+A_1I(I+1) \enspace \makebox{for $I$ even}
\end{eqnarray}
Considering $E_1$, $A_1$ and $B$ as parameters,  we obtained from the  energies of three levels
$E_1$ = 1127.3 keV, $A_1 = 7.63$ keV and $B = -0.47$ keV. This corresponds to a moment of inertia of
65.5 MeV$^{-1}$ (see Fig.~\ref{fig:moments}). The difference to the moment of inertia of the $0^-$ band
is quite large and the energy of 1173.1 keV of the 2$^-$ level of thus assumed $1^-$ band is much higher than
the energy of 1016.8 keV of the 2$^-$ level of the assumed the 2$^-$ band (should be the opposite).
If, however, we consider the  1173.1 keV state as the bandhead of the $K^{\pi}=2^-$ band, then the moment
of inertia is determined as 78.5 MeV$^{-1}$ close to the moment of inertia of the $K^{\pi}=0^-$ band.
Although  some ambiguity remains, the level  3$^-$ at  1264.8 keV  can be proposed as  the bandhead
of the $K^{\pi} = 3^-$ band.

In a more advanced model \cite{Sol76}, that takes into account the Coriolis
interaction between all octupole bands, one can fit 11 parameters
(bandhead energies, rotation parameters and Coriolis intrinsic
matrix elements between bandheads) to the experimental energies.
The former level
assignment gives $E_0 = 551.0$ keV, $E_1 = 1136$ keV, $E_2 = 1006$ keV,
$E_3 = 1241$ keV with $\chi^2 = 0.97$,
while the latter gives a slightly better value of $\chi^2 = 0.63$ and
resonable values of the fitted energies
$E_0 = 551.6$ keV, $E_1 = 987.4$ keV, $E_2 = 1158$ keV,
$E_3 =1240$ keV. However, the predicted $1^-$ bandhead at 994 keV is not
observed experimentally.

\textsl{The 2$^+$, 4$^+$ and 6$^+$ states.}
States with spins and parities firmly assigned as 2$^+$ excitations
 dominate in the triton spectra. Some of them are assigned as members of $K^{\pi}=0^+$ bands
 and others  probably are bandheads and the levels with spins 4$^+$ and even 6$^+$ are identified
 as possible members of these bands. From the analysis of the $\gamma$-spectra \cite{Bro06}
the 866.8 keV state was identified as the bandhead of the $\gamma$-vibrational band with
members  911.5 and 970.7 keV. A possible continuation of this band can be the 1132.7 keV state,
since for this state the typical  2$^+$ angular distribution is distorted by a possible
admixture of a 6$^+$ state (it might be a doublet of  2$^+$ and 6$^+$ states). The 1132.7 keV
state was tentatively  identified as the bandhead of the  $K^{\pi}=2^+$ in \cite{Wei72} using
the analysis of the $\gamma$ spectra. The  4$^+$ states at 1226.8 keV
and the 6$^+$ state at 1372.0 keV  could be proposed as  members of this band. However, the cross section
for the 4$^+$ state  exceeds the one for the 2$^+$ state, contrary to the above conditions.
Therefore, the $K^{\pi} = 4^+$ band is offered as an option.

\textsl{The 0$^+$ states.}
For the state at the energy of 927.3 keV assigned in \cite{Ard94} and
in this study as a 0$^+$ state no members of the band were observed in the (p,t) reaction.
The energy 967.6 keV  was accepted for the 2$^+$ member of the band as suggested  in \cite{Ard94}.
The 0$^+$ state at 1277.2 keV is strongly
 excited and the members of the assumed band  have to be excited too. A clear sequence
 of states is observed with a spin assignment of2$^+$, 4$^+$ and 6$^+$,   as can be seen
 in Table~\ref{tab:bands}, but moments of inertia determined from this sequence are very high
 (124 MeV$^{-1}$ from the 2$^+$ and 0$^+$ state energy difference) and are decreasing as a function of spin.
 Other  possible sequences  do not meet the conditions set forth above.
 A possible
 explanation for such a behavior  was suggested in \cite{Lev13} but it is hardly applicable
 in this case.  An assumption may be suggested
 that the structure of this state is different from other collective states.
 To some extent the same remark
 can be attributed to the band probably built on the 0$^+$ state at 1569.0 keV,
 whose moment of inertia weakly decreases as a function of spin, though weakly.
The 0$^+$ states at 1797.0, 1821.8  and 1861.0  keV are  also strongly excited and
 the excitation of other members of the assumed bands have to be seen in the (p,t) reaction.
 At least the 2$^+$ members  can be attributed to such bands built on the 0$^+$ states at
 1797.0 and  1821.8 keV. For the 0$^+$ state at 1861.0  keV, no prolongation of the band
 is clearly visible in the triton spectra.
 Only the unlikely assumption can be made that the corresponding line is hidden under the 1915.2 keV line,
  but the moment of inertia of 55.4 MeV$^{-1}$ from this assumption is much less than the one for the ground state.
 An ambiguous situation is met also for the $0^+$ state at 1931.0 keV,  whose excitation  is
 only  slightly weaker than the first excited state at 691.4 keV.
 Two different sequences may be assumed for the band built on this state, but for both the moment of
 inertia is decreasing with spin.
 As it was noted already, the angular distribution for the state at 2917.4 keV differs considerably from
all others  and can be fitted only by the calculation for transfer of the  $(1j_{15/2})^2$
neutron configuration.

\newcolumntype{d}{D{.}{.}{3}}
\begin{table*}[]
\caption{\label{tab:bands} \normalsize{The sequences of  states
which can belong to rotationak bands (from  the CHUCK fit, the (p,t) cross
sections and the inertial parameters). More accurate values of
energies are taken from the first two columns of
Tab.~\ref{tab:expEI}.}}
\begin{ruledtabular}
\begin{tabular}{ccccccccc}\\
 $0^+$ & $1^+$ & $2^+$ & $3^+$ & $4^+$ & $5^+$ & $6^+$ & $7^+$
& $ 8^+$ \\
\hline\\
0.0 &&    47.6 &&     156.6 &&   322.7 &&   541.1 \\
691.4 && 734.6 && 833.1 && 984.9 && 1186.6 \\
    &&   866.8 & 911.5 & 970.7 &  & (1132.7) &  &   \\
    &&   1132.7             &&1226.8   && 1372.0 &&\\
or        &&                &&1226.8   && 1372.0 &&\\
    &&                & &1194.0  &  &1314.8  &  &  \\
927.3&& 967.6 &&&&&&\\
1277.2&& 1301.4 && 1361.5 && 1460.4&&\\
      && 1489.2 && 1572.9 && 1700.5 &&\\
1482.2 && 1520.4 &&1605.4&&1737.4&&\\
1569.0 && 1600.2 && 1673.2 && 1790.8 &&\\
       && 1647.7 && 1744.4 && 1880.8 &&\\
1797.0 && 1838.6 &&  && &&\\
1821.8 && 1870.9 &&&&&&\\
1861.0 && (1915.2) &&  &&&&\\
1931.0 && 1970.7 &&2068.6&&  &&\\
or &&  && &&  &&\\
1931.0 && 1977.8 &&&&  &&\\
     &&2059.8 && 2135.9 && 2254.4 &&\\
     &&2146.5 && 2231.3 &&&&\\
     &&2203.8 && 2312.2 &&&&\\
     &&2418.8 && 2527.2 &&&&\\
     && 2673.5 && 2754.3 && 2878.3 &&\\
     && 2779.1 && 2889.8 && 3058.3 &&\\
     && 2791.0 && 2899.2 &&&&\\
2917.4 && 2959.7 &&&&&&\\
\hline\\
& $1^-$ & $2^-$ & $3^-$ & $4^-$ & $5^-$ & $6^-$ & $7^-$ & $8^-$ \\
\hline\\
& 563.2 && 629.0 && 746.8 && 915.2 &\\
       && 1016.8 & 1050.9 & 1098.2 & 1155.4 &&&\\
& (1141.5) &1173.06& 1211.3 &&(1321.8)  &&  &\\
&&&1264.8&&1391.7 &&&\\
&&&1679.8&&1758.9 &&&\\
\end{tabular}
\end{ruledtabular}
\end{table*}

\textsl{Moments of inertia (MoI).}
The moment of inertia of the g.s. of $^{232}$U is 63 MeV$^{-1}$, and as such much higher than in $^{228}$Th
and $^{230}$Th.  The $K^{\pi}= 0^-$ band with the bandhead
1$^-$ at 563.2 keV is well established, it exhibits an MoI  increase of about 20\,\% and can serve as orientation
for such excitations. The assumption that the 2$^-$ level at 1173.1 keV belongs to the $K^{\pi}= 1^-$ band
is not confirmed by the present analysis (see above).
If, however, we assume that the state at 1173.1 keV is the bandhead of
the $K^{\pi}=2^-$ band, then the MoI determined from the $3^{-} - 2^{-}$
energy difference is 78.5 MeV$^{-1}$,  only slightly higher than the MoI
of the $K^{\pi}=0^-$ band.
Although  the ambiguity remains, the level 2$^-$ at 1016.9 keV is the member
of the $K^{\pi}= 1^-$ band
(with the 1$^-$ level not observed) and  the   3$^-$ at level 1264.8 keV
can be proposed as  the bandhead
of the $K^{\pi} = 3^-$ band.

Unlike in $^{228}$Th and $^{230}$Th, the MoI of the bands built on the excited 0$^+$ states
in $^{232}$U are not much higher than  those for the g.s. band. Only two bands, starting
at 1227.2 and 1569.0 keV, have a significant excess of the MoI. The principal difference
is that the first excited 0$^+$ state in $^{232}$U seems to be a $\beta$-vibrational state.  The results of
the analysis of $\gamma$ spectra and the MoI value close to the one of the g.s.  give evidence for such
conclusion. At the same time  the first excited $0^+$ state in $^{228}$Th  \cite{Lev13}
 as well as in $^{240}$Pu \cite{Spi13} may have an octupole two-phonon nature. The first
 excited (possible $\beta$-vibrational) state in $^{232}$U is  most strongly excited,   just the same  as the
 the first excited (octupole two-phonon) states in $^{228}$Th and $^{240}$Pu and the first excited state
 in  $^{230}$Th with a more complicated phonon structure \cite{Lev09}.
In $^{232}$U, the state at 927.3 keV was suggested in \cite{Ard94} to be an octupole two-phonon excitation.
It is confirmed by the large values of the  $B(E1)/B(E2)$ ratio calculated
using the data on the $\gamma$-intensities  for transitions from $0^+$ and $2^+$ levels of this band to
the levels of the 0$^-$ and g.s. bands \cite{Ard94} (see discussion below and Tables \ref{BE2/BE1_IBM} and
\ref{BE_SQPM}).

The MoI of the 927.3 keV band confirms this assignment, though it is only 16\,\% larger than that of the g.s.,
compared to a larger excess of 36\,\% and 23\,\% for $^{228}$Th and $^{230}$Th, respectively.
All these facts indicate that the strong population of the first excited 0$^+$ states does not allow to
identify their  phonon structure.

The value of the MoI for the band built on the  2$^+$  state at 866.8 keV is close to the one
for the possible $\beta$-vibration band (they are only 6\,\% and 8\,\% larger than those of the g.s. band),
thus confirming its interpretation as $\gamma$-vibrational band.
As for the experimental evidence of the nature of other $0^+$ states, we derived only values of the MoI
  from the sequences of states treated as rotational bands and thus only
tentative conclusions can be drawn about their structure. In contrast to $^{230}$Th
\cite{Lev09},  for which they are distributed almost uniformly over the region
until 1.80 of the g.s. value (equal to 56 MeV$^{-1}$) and to $^{228}$Th, for which most of values are
 in the range of 1.35 - 1.85 of the g.s. value (equal to 52 MeV$^{-1}$), most of the MoI values of
 the $0^+$ states in  $^{232}$U do not exceed the value of 1.27 of the g.s. (equal to 63 MeV$^{-1}$).
 This fact can  indicate that  the corresponding states are possibly of quadrupole one-phonon or two-phonon nature.

\subsection{\label{sec:IBM} IBM calculations}

The  Interacting Boson Model (IBM) describes the low-lying positive and negative parity states by
treating the valence nucleons in pairs as bosons. The positive parity states are described by
introducing \(s\) and \(d\) bosons, which carry angular momentum of \(L^{\pi}\)=0\(^{+}\)
 and \(L^{\pi}\)=2\(^{+}\), respectively, while the negative parity states can be calculated
  by additionally including \(p\) and \(f\) bosons having \(L^{\pi}\)=1\(^{-}\) and
   \(L^{\pi}\)=3\(^{-}\), respectively. In the present paper the IBM-1 version of the model
   is used, which means that no distinction is made between protons and neutrons \cite{Iach87}.
   Full IBM-\(spdf\) calculations have been previously done with success in Refs. \cite{Eng87,Zam01,Zam03}.

The octupole degree of freedom is well known for playing a major role in the actinide region
\cite{Rob12,Rob13}. In fact, octupole correlations have been predicted to be present in the
$Z \sim 88$ and $N \sim 134$ region
\cite{Naz84} and have attracted a lot of experimental investigations
 centered on energy spectra and transition probabilities \cite{But96}. The low-lying properties
 of these nuclei have been interpreted using a series of theoretical models, including
 the IBM \cite{Zam01,Zam03}, which mainly concentrated on the study of electromagnetic decay properties.
 In the last years, several nuclei in this region were investigated using the (p,t) reaction and
 the  transfer intensities became available also \cite{Wir04,Spi13}. Therefore, several models
 tried to describe the complete experimental situation \cite{Lev13,Spi13,LoI05}.

As presented also in this paper, an increased number of 0\(^+\) excitations have been
populated in the previous two-neutron transfer experiments with a rather high
intensity \cite{Wir04,Lev09,Lev13,Spi13}. Since some of these states strongly decay to
the negative parity states, it is believed that the quadrupole and octupole degrees of
freedom are closely connected to these excitations. In the IBM, such 0\(^+\) states have
been interpreted as having a double octupole character \cite{Lev13,Spi13}. Although this simple
picture may not be entirely correct, the IBM has been proved to reasonably describe simultaneously
the electromagnetic and transfer  properties. In order to reproduce the experimental features,
one has to abandon the description of the nuclei using a simplified Hamiltonian, which is suited
to describe mainly electromagnetic data. Such calculations were found to completely fail to
reproduce the (p,t) spectroscopic factors by predicting a transfer strength of 1\(\%\) of that
of the ground state, while experimentally the summed transfer intensity amounts to about 80\(\%\)
in this region. The solution seems to be the introduction of the second-order O(5) Casimir
operator in the Hamiltonian, which allows for a far better description of the complete experimental data.

In the present work, calculations were performed in the \(spdf\) IBM-1 framework using the Extended
Consistent Q-formalism (ECQF) \cite{Cas88}. The Hamiltonian employed  in the present paper is:

\begin{eqnarray}
\hat{H}_{spdf}=\mathrm{\epsilon}_{d} \hat{n}_{d}+\mathrm{\epsilon}_{p}
\hat{n}_{p}+\mathrm{\epsilon}_{f} \hat{n}_{f} + \mathrm{\kappa}(\hat{Q}_{spdf}\cdot
\hat{Q}_{spdf})^{(0)}\nonumber\\
 + \mathrm{\mathit a_{3}}
[(\hat{d}^{\dagger}\tilde{d})^{(3)} \times (\hat{d}^{\dagger}\tilde{d})^{(3)}]^{(0)}\label{eq1}
\end{eqnarray}
where \(\epsilon_{d}\), \(\epsilon_{p}\), and \(\epsilon_{f}\) are the boson energies
and \(\hat{n}_{p}\), \(\hat{n}_{d}\), and \(\hat{n}_{f}\) are the boson number operators.
In the \(spdf\) model, the quadrupole operator is considered as being \cite{Kuz90}:
\begin{eqnarray}
\hat{Q}_{spdf}=\hat{Q}_{sd}+\hat{Q}_{pf}=\nonumber\\
(\hat{s}^{\dagger}\tilde{d}+\hat{d}^{\dagger}\hat{s})^{(2)}+\chi^{(2)}_{sd}(\hat{d}^{\dagger}\tilde{d})^{(2)}
+\frac{3\sqrt{7}}{5}[(p^{\dagger}\tilde{f}+f^{\dagger}\tilde{p})]^{(2)}\nonumber\\
+\chi^{(2)}_{pf} \left\{ \frac{9\sqrt{3}}{10}(p^{\dagger}\tilde{p})^{(2)}
+\frac{3\sqrt{42}}{10}(f^{\dagger}\tilde{f})^{(2)}\right\}\label{eq2}
\end{eqnarray}

The quadrupole electromagnetic transition operator is:
\begin{eqnarray}
\hat{T}(E2)=e_{2} \hat{Q}_{spdf}\label{eq3}
\end{eqnarray}
where \(e_{2}\) represents the boson effective charge.

The \(E1\) transitions are described in the IBM by a linear combination of the three allowed one-body interactions:
\begin{eqnarray}
\hat{T}(E1)=e_{1}[\chi_{sp}^{(1)}({s}^{\dagger}\tilde{p}+{p}^{\dagger}\tilde{s})^{(1)}+({p}^{\dagger}\tilde{d}
+{d}^{\dagger}\tilde{p})^{(1)}\nonumber\\
+\chi_{df}^{(1)}({d}^{\dagger}\tilde{f}+{f}^{\dagger}\tilde{d})^{(1)}]\label{eq4}
\end{eqnarray}
where \(e_{1}\) is the effective charge for the \(E1\) transitions and \(\chi_{sp}^{(1)}\)
and \(\chi_{df}^{(1)}\) are two model parameters.

At this point, one has to introduce an additional term in order to describe the connection between
states with no (\(pf\)) content with those having \((pf)^2\)  components. This term is very useful
to describe both the \(E2\) transitions and also the transfer strength between such states.
Therefore, the same dipole-dipole interaction term is introduced in the present calculations
as previously used in Refs. \cite{Eng87,Zam03,Lev13}:
\begin{eqnarray}
\hat{H}_{int}=\alpha \hat{D}^{\dagger}_{spdf}\cdot \hat{D}_{spdf}+ H.c.\label{eq6}
\end{eqnarray}
where
\begin{eqnarray}
\hat{D}_{spdf}=-2\sqrt{2}[{p}^{\dagger}\tilde{d}+{d}^{\dagger}\tilde{p}]^{(1)}
+\sqrt{5}[{s}^{\dagger}\tilde{p}+{p}^{\dagger}\tilde{s}]^{(1)}\\\nonumber
+\sqrt{7}[{d}^{\dagger}\tilde{f}+{f}^{\dagger}\tilde{d}]^{(1)}\label{eq7}
\end{eqnarray}
is the dipole operator arising from the \(O\)(4) dynamical symmetry limit, which does not conserve
separately the number of positive and negative parity bosons \cite{Kuz89,Kuz_up}.

\begin{figure*}
\epsfig{file=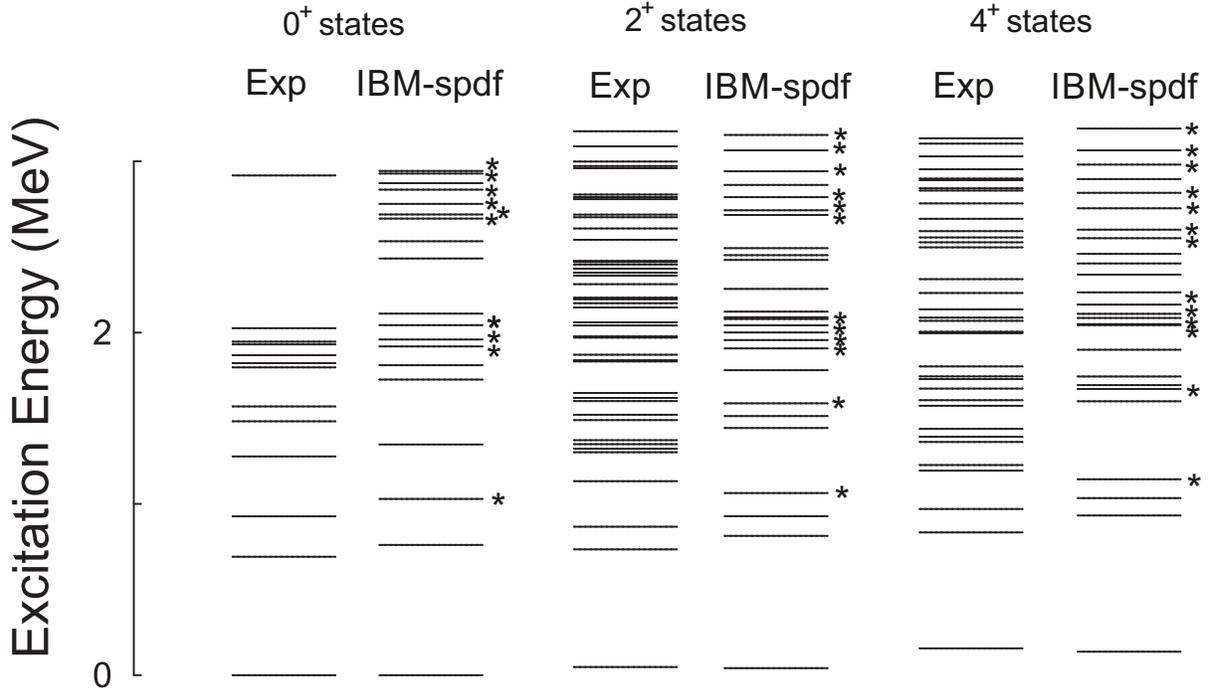, width=16 cm, angle=0} \caption{\label{fig:IBMspe} Energies of all
experimentally assigned excited 0\(^{+}\), 2\(^{+}\), and 4\(^{+}\) states in \(^{232}\)U in
comparison with IBM-\(spdf\) calculations. The states containing 2 \(pf\) bosons in their
structure and assumed to have a double dipole/octupole character are marked with an asterisk.}
\end{figure*}

The goal of the present paper is to describe simultaneously both the existing electromagnetic
and the transfer strength data. To achieve this goal, two-neutron transfer intensities between
the ground state of the target nucleus and the excited states of the residual nucleus were also
calculated. The $L=0$ transfer operator has the following form in the IBM:

\begin{eqnarray}
\hat{P}^{(0)}_{\nu}=(\alpha_{p}\hat{n}_{p}+\alpha_{f}\hat{n}_{f})\hat{s}+\nonumber\\
+\alpha_{\nu} \left (\Omega_{\nu}-N_{\nu}-\frac{N_{\nu}}{N} \hat{n}_{d}\right)^{\frac{1}{2}}
\left(\frac{N_{\nu}+1}{N+1}\right)^{\frac{1}{2}} \hat{s}\label{eq5}
\end{eqnarray}
where \(\Omega_{\nu}\) is the pair degeneracy of the neutron shell, \(N_{\nu}\) is the number
of neutron pairs, \(N\) is the total number of bosons, and \(\alpha_{p}\), \(\alpha_{f}\),
and \(\alpha_{\nu}\) are constant parameters. In this configuration, the $L=0$ transfer operator
contains additional terms besides the  leading order term (third term) \cite{Iach87}, which ensures
a non-vanishing transfer intensity to the states with \((pf)^2\) configuration.

The calculations were performed using the computer code OCTUPOLE \cite{Kuz_up} by allowing up to
three negative parity bosons. The following parameters in the Hamiltonian have been used:
\(\epsilon_{d}\)=0.27 MeV, \(\epsilon_{p}\)=1.14 MeV, \(\epsilon_{f}\)=0.95 MeV, \(\kappa\)=-13 keV,
and \(a_{3}\)=0.026 MeV, which ensures a good reproduction of the low-energy states.
The interaction strength is given by the \(\alpha\) parameter and is chosen to have a very small value,
\(\alpha\)=0.0005 MeV, similar to Refs. \cite{Zam01,Zam03}, which has a very small influence on the level energies.

The most important result of these (p,t) transfer experiments is the fact they reveal a large number of 0\(^+\) states, the presence of such states at higher excitation energies being the subject of intensive theoretical investigations. Therefore, we present in Fig. \ref{fig:IBMspe} the full spectrum of experimental excited 0\(^+\) states in comparison with the corresponding calculated values. The IBM predicts the existence of 19 0\(^+\) states up to an excitation energy of 3 MeV in comparison with 13 0\(^+\) states excited in the experiment in the same energy range. The calculated distribution of 0\(^+\) states is very similar to the experimental one up to around 2 MeV. The situation is completely different between 2 and 3 MeV, where a large gap is seen in the experiment up to 2.9 MeV, while the IBM predicts an increased number of states with increasing excitation energy. In the experiment, we can speculate that in this region the 0\(^+\) states  carry very small transfer strengths, therefore the sensitivity of our experiment was not sufficient to discriminate between individual states. Most of the calculated excitations in this energy range are having two \(pf\) bosons in their structure (states marked with asterisk), therefore being related to the presence of double dipole/octupole structure \cite{Zam01}. Although it is very interesting that IBM describes both the electromagnetic and transfer data at the same time, this is most likely not the only mechanism providing an increased number of 0\(^+\) states and therefore we cannot make a definite conclusion on the nature of these excitations based only on these limited experimental data. To support such a claim, more experimental information is needed and in particular the \(E1\) and \(E3\) transition probabilities to the negative parity states. In Fig. \ref{fig:IBMspe}, the 2\(^{+}\) and 4\(^{+}\) levels revealed in the present experiment are also compared with the predictions of the IBM. The experiment revealed 40 firmly assigned excited 2\(^{+}\) states and 26 solid assigned excited 4\(^{+}\) states up to 3.2 MeV. In the same energy range, the calculations produced 26 excited 2\(^{+}\) states and 26 4\(^{+}\) excitations.

Since the octupole degree of freedom plays an important role in this mass region, it is crucial for a model to describe at the same time at least the B(E1) and the B(E1)/B(E2) ratios, if the reduced transition probabilities have not been measured. In the IBM, the \(E1\) transitions are calculated with Eq.(\ref{eq4}), while for the \(E2\) transitions using Eq.(\ref{eq3}). In the present calculations, we have used (\(\chi^{(2)}_{sd}\)=-1.32, \(\chi^{(2)}_{pf}\)=-1) as the quadrupole operator parameters and \(\chi_{sp}\)=\(\chi_{df}\)=-0.77 for the parameters in Eq.(\ref{eq4}). The remaining parameters are the effective charges and are used to set the scale of the corresponding transitions: \(e_{1}\)=0.0065 \(e\)fm and \(e_{2}\)=0.184 \(e\)b.

\begin{table}
\caption{\label{BE2/BE1_IBM} Experimental and calculated  B(E1)/B(E1) (from  the 0$^{-}_{1}$ state) and   B(E1)/B(E2) (from the 0$^{+}_{3}$ state)  transitions ratios in $^{232}$U. The parameters
of the \(E1\) operator are fitted to the experimental data available. The B(E1)/B(E2) ratios are given in units
of 10$^{-4}$ b$^{-1}$.}
\begin{ruledtabular}
\begin{tabular}{ccccccc}\\
K$^{\pi}$ & E$_{i}$ (keV) & J$_{i}$ & J$_{f1}$ & J$_{f2}$ & Exp. & IBM \\
\hline\\
0$^{-}_{1}$   & 563   & 1$^{-}$ & 2$^{+}_{1}$ & 0$^{+}_{1}$   & 1.89(8)    & 1.89     \\
                      & 629   & 3$^{-}$ & 4$^{+}_{1}$ & 2$^{+}_{1}$   & 1.19(5)    & 1.17     \\
                      & 747   & 5$^{-}$ & 6$^{+}_{1}$ & 4$^{+}_{1}$   & 0.94(8)    & 0.97     \\
0$^{+}_{3}$  & 927   & 0$^{+}$ & 1$^{-}_{1}$ & 2$^{+}_{1}$   & 44(7)       & 58     \\
                      & 968   & 2$^{+}$ & 1$^{-}_{1}$ & 0$^{+}_{1}$   & 150(30)    & 122     \\
                      &         & 2$^{+}$ & 1$^{-}_{1}$ & 2$^{+}_{1}$   & 45(1)       & 85     \\
                      &         & 2$^{+}$ & 1$^{-}_{1}$ & 4$^{+}_{1}$   & 24(1)       & 47   \\
                      &         & 2$^{+}$ & 3$^{-}_{1}$ & 0$^{+}_{1}$   & 337(68)    & 167      \\
                      &         & 2$^{+}$ & 3$^{-}_{1}$ & 2$^{+}_{1}$   & 101(3)     & 117   \\
                      &         & 2$^{+}$ & 3$^{-}_{1}$ & 4$^{+}_{1}$   & 54(1)       & 65   \\
\end{tabular}
\end{ruledtabular}
\end{table}

The B(E1)/B(E2) ratios discussed in Table \ref{BE2/BE1_IBM} belong to the K\(^{\pi}\)=0\(_{3}^{+}\) band (the predicted double octupole phonon band). All the states belonging to this band are having \((pf)^2\) bosons in their structure in the IBM calculations and are supposed to have a double octupole phonon character. The agreement in Table \ref{BE2/BE1_IBM} between experiment and calculations is remarkably good, giving even more confidence in the structure proposed by the IBM. If other excited 0\(^+\) levels decay to the negative parity states, one would need the crucial information about the decay pattern of these levels. This can be achieved by future (p,t\(\gamma\)) and (n,n'\(\gamma\)) experiments and we stress here the necessity of performing such delicate investigations.

\begin{figure}
\begin{center}
\epsfig{file=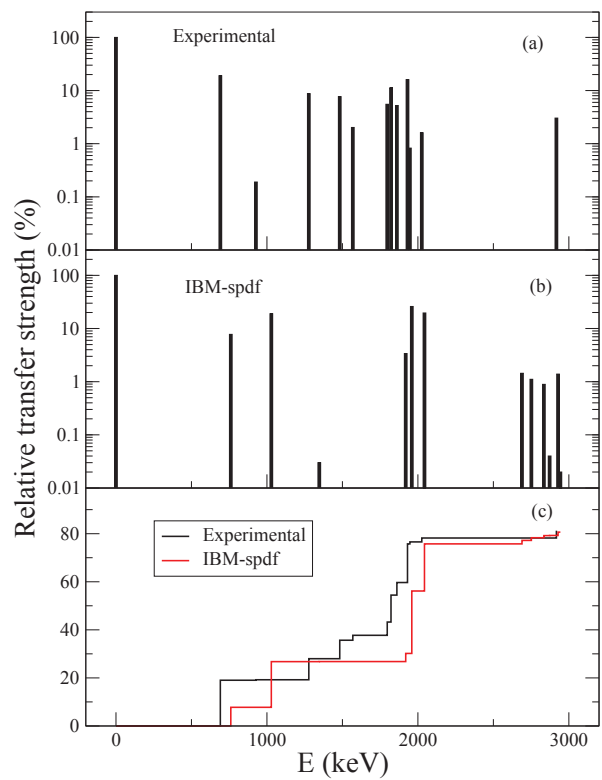, width=8 cm, angle=0} \caption{\label{fig:transfer} (Color online) Comparison between the experimental (both firm and tentatively assigned states are included in the figure) two-neutron transfer intensities [panel (a)] for the 0\(^{+}\) states and the IBM predictions [panel (b)]. In panel (c) the experimental versus computed running sum of the (p,t) strengths is given.}
\end{center}
\end{figure}

The experimental integrated two-neutron transfer intensities are displayed in Fig. \ref{fig:transfer} panel (a). In contrast to \(^{228,230}\)Th where the spectrum is dominated by a single state with high cross section of about 15-20\(\%\) of that of the ground state, the transfer intensity in \(^{232}\)U goes not only to the first excited 0\(^+\) state, but also to a group of states around 2 MeV, which carries more than 30\(\%\). In the IBM [Fig. \ref{fig:transfer} panel(b)], the transfer intensity is also split between the first two excited 0\(^+\) states and a group of 0\(^+\) excitations around 2 MeV. To better compare the agreement with the experimental data, one has to look also at the summed transfer intensity, which is presented in Fig. \ref{fig:transfer} panel (c) for both the experimental and the calculated values. The main characteristics of the experimental transfer distribution are reproduced, namely the increased population of two groups of 0\(^+\) excitations around 1 and 2 MeV, but IBM fails to give a detailed explanation of the individual states. To perform the IBM calculations, the parameters from Eq.(\ref{eq5}) were estimated from the fit of the known two-neutron transfer intensities (integrated cross sections) in Table 1.  The values employed in the present paper are \(\alpha_{p}\)=0.51 mb/sr, \(\alpha_{f}\)=-0.45 mb/sr, and \(\alpha_{1}\)=0.013 mb/sr.

\subsection{\label{sec:QPM} QPM calculations}

To obtain a detailed information on the properties of the states excited in the (p,t) reaction, a microscopic approach is
necessary. The ability of the QPM to describe multiple $0^+$ states (energies, $E2$ and $E0$ strengths, two-nucleon spectroscopic factors) was
demonstrated for $^{158}$Gd \cite{LoI04}. In a subseqent paper the QPM was applied to the microscopic structure of $0^+$ states in $^{168}$Er and three actinide nuclei ($^{228}$Th,
$^{230}$Th and $^{232}$U) \cite{LoI05}. Single particle basis states up to 5 MeV were generated by a deformed axially-symmetric  Woods-Saxon potential. Two-body potentials were represented by a monopole plus multipole pairing interaction and isoscalar and isovector multipole-multipole interactions. Two-phonon states were calculated for multipolarities $\lambda = 2 - 5$. These calculations are also used to compare to the present detailed analysis of the experimental data for $^{232}$U. As
for the theoretical basis of the calculations, we refer to the publications \cite{LoI05,Sol92}.

The (p,t) normalized relative transfer
spectroscopic strengths in the QPM are expressed as ratios
\begin{equation}\label{spec-fact}
S_n(p,t) = \left[\frac{\Gamma_n(p,t)}{\Gamma_0(p,t)}\right]^2 \ ,
\end{equation}
where the amplitude $\Gamma_n(p,t)$ includes the transitions between the $^{234}$U ground state and one-quadrupole $K=0$ phonon components of the $^{232}$U wave function. The amplitude $\Gamma_0(p,t)$ refers to the transition between the $^{234}$U and $^{232}$U ground states. The selected normalization assures that $S_0(p,t) = 100$ for the ground state transition.

\begin{figure}
\centering \epsfig{file=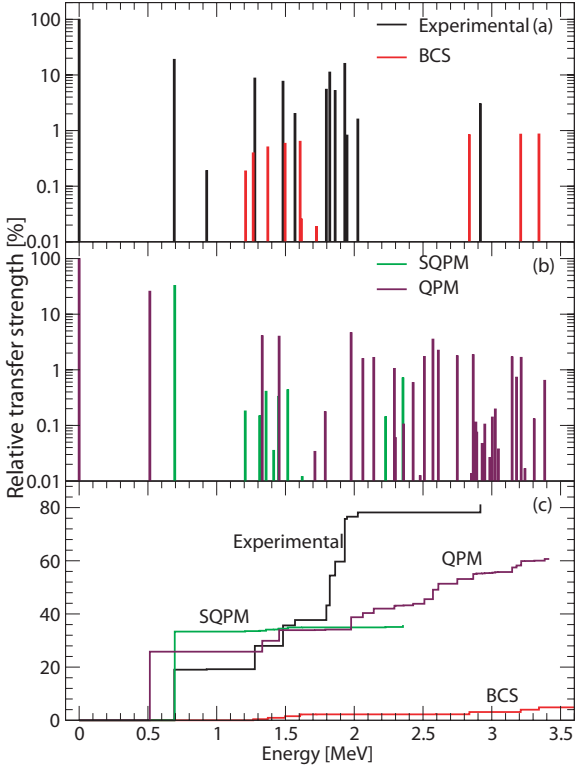, width=8 cm, angle=0} \caption{
\label{fig:QPM-BCS}
(Color online) Comparison of experimental and BCS in panel (a), SQPM and QPM in panel (b) 0$^+$ (p,t) normalized relative  strengths. The value for the 0$^+_{g.s.}$ is normalized to 100.
    The experimental increments of the (p,t) strength in comparison to the QPM, SQPM and BCS calculations are shown in the lower panel (c).}
\end{figure}

To see the role of the two-phonon and pairing-vibrational excitations in the QPM calculations, we performed  simple QPM (SQPM) calculation using the Nilsson potential plus monopole pairing
interaction (Nilsson parameters $\kappa$ and $\mu$ taken from Ref.~\cite{Sol76}, deformation
parameters $\epsilon_2 =0.192$, $\epsilon_4=-0.008$ and pairing gaps $\Delta_p =0.706$ MeV,  $\Delta_n=0.602$ MeV for $^{232}$U and
$\epsilon_2 =0.200$, $\epsilon_4=-0.073$ and pairing gaps $\Delta_p =0.738$ MeV, $\Delta_n=0.582$ MeV for $^{234}$U from
Refs.~\cite{Mol95,Mol97}) plus isoscalar and isovector quadrupole-quadrupole and octupole-octupole interactions. Only
one-phonon RPA states were taken into account  in these calculations. Energies of two-quasiparticle $0^+$ states were estimated from the BCS theory.
The model predicts 15 neutron two-quasiparticle states of the structure $\alpha^{\dag}_q \alpha^{\dag}_{\bar{q}}$ below 4 MeV that correspond to broken neutron pairs
sensitive to two-neutron transfer. The energies and normalized relative transfer strengths are shown in Fig.~\ref{fig:QPM-BCS}(a) for $S_n(p,t) \ge 0.01$ and compared to the experimental energies and relative transfer strengths. It is evident that the two-quasiparticle $0^+$ states represent only a minor contribution to the total relative transfer strength (cf. Fig.~\ref{fig:QPM-BCS}(c)).

\begin{figure}
\begin{center}
\epsfig{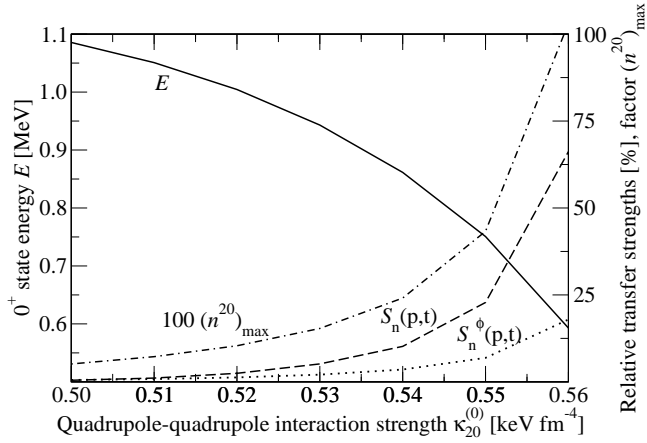}
\caption{\label{fig:lowest0+} The SQPM energy $E$ (solid line), the normalized relative transfer strength $S_n(p,t)$ (dashed line), the contribution $S^{\phi}_n(p,t)$ of
the backward RPA amplitude $\phi$ to $S_n(p,t)$ (dotted line) and the maximum number of quasiparticles with quantum number $q$ in the ground state
$(n^{20})_{max}$ (dashed-dotted line) as
functions of the isoscalar quadrupole-quadrupole interaction strength $\kappa^{(0)}_{20}$ for the first excited $0^+$ state, $\kappa^{(1)}_{20} = 0$.}
\end{center}
\end{figure}

The strengths of the isocalar and isovector
quadrupole-quadrupole interactions in the SQPM, $\kappa^{(0)}_{20}$ and $\kappa^{(1)}_{20}$, respectively, were varied to fit the experimental
energies and (p,t) spectroscopic strengths of the lowest $0^+$ states. It was found that an effect of $\kappa^{(1)}_{20}$ on the (p,t) spectroscopic strengths is negligible and that
$\kappa^{(0)}_{20}$ significantly influences only energy and spectroscopic strength of the first $0^+$ excited state.
It is known that in the even-even actinides the phonon coupling does not spoil the coherence of pairing correlations in the lowest $0^+$ excited state \cite{LoI05}. As a consequence, the state has a pronounced pairing-vibrational character that manifests itself by large RPA backward $\phi$ amplitudes. From Fig.~\ref{fig:lowest0+} one can see that the contribution $S^{\phi}_n$(p,t) of the backward RPA amplitudes to the normalized relative transfer spectroscopic strength $S_n$(p,t) is important for the first excited $0^+$ state, thus indicating its pairing vibrational character. The pairing interaction is essential for reproducing the experimental relative transfer strength of the first excited $0^+$ state. If we artificially lower the neutron and proton pairing interaction strengths and simultaneously change the isoscalar quadrupole-quadrupole interaction to fit the experimental energy of the first excited $0^+$ state both $S^{\phi}_n$(p,t) and $S_n$(p,t) rapidly drop down. The SQPM predicts $B(E2)= 4$ W.u. for the transition from the first excited $0^+$ state to the $2^+$ member of the g.s. band, the QPM gives a slightly lower value of 2.3 W.u. Therefore, we can assume that the lowest $0^+$ excited state ($0_2^+$) has a mixed pairing-vibrational and $\beta$-vibrational character. The contribution of $S^{\phi}_n$(p,t) for higher excited $0^+$ states in the SQPM is significantly lower and in most cases negligible, thus indicating their
weak phonon-vibrational or two-quasiparticle character. The maximum value of the number of quasiparticles with the quantum number $q$ in
the ground state, $n^{20}_{max}$, measures the ground-state correlations and can be calculated from (see \cite{Web98}):
\begin{equation}
n^{20}_{max} = \max{\left[ \frac{1}{2}(\phi^{20}_{qq})^2 \right] }
\end{equation}
where $\phi^{20}_{qq}$ are the backward RPA amplitudes of the first $0^+$ excited state.
For the isoscalar quadrupole-quadrupole interaction strength $\kappa^{(0)}_{20} = 0.554$ keV fm$^{-4}$, that reproduces the experimental energy of the first $0^+$ excited
state, the ground-state correlations estimated by $n^{20}_{max}$ become large (see Fig.~\ref{fig:lowest0+}). As a consequence, the RPA approximation used in the SQPM is no more accurate and multi-phonon admixtures and interactions between phonons have to be taken into account.

In Fig.~\ref{fig:QPM-BCS} (a) and (b) the experimental spectrum of the $0^{+}$ (p,t) normalized relative transfer strengths is compared to the results of the SQPM and QPM calculations. The numerical results of the QPM calculations from Ref. ~\cite{LoI05} are provided to us by A.~V.~Sushkov \cite{Sush15}.
Both SQPM and QPM calculations reproduce the strong excitation of the first 0$^+$ excited state in accordance with the experiment. The SQPM generates 9 $0^+$ states below 2 MeV in
fair agreement with the 10 firmly assigned states and 3 $0^+$ states in the region $2-3$ MeV compared to 2 experimentally assigned states. The QPM fails to reproduce the experimental number of the $0^+$ states. It predicts only 6 $0^+$ states below 2 MeV and 20 $0^+$ states in the region $2-3$ MeV. The difference in the number of the $0^+$ states between the SQPM
and the QPM is caused mainly by the truncated SQPM model space (two-phonon states not considered).

In Fig.~\ref{fig:QPM-BCS} (c) we present also the increments of the experimental relative transfer strength in comparison to those of the BCS, SQPM and QPM. Additional interactions
in the QPM lead to level repulsion (excited $0^+$ states spectrum broadening) and transfer strength fragmentation (lower relative transfer strength for the first excited
$0^+$ state in favor of higher excited states up to 2 MeV). In the region above 1.8 MeV, both SQPM and QPM fail to reproduce the sharp experimental increase of the (p,t) strength
running sum.
In Table~\ref{QPM_structure}, structure and normalized relative transfer strength of the QPM $0^+$ excited states are
presented. It is difficult to make an assignment to experimental levels above 1.5 MeV. The second two excited states, $0_3^{+}$ and $0_4^{+}$, most probably correspond to the experimental levels at 1.277 and 1.482 MeV, which is supported by the similar normalized relative transfer strengths. The experimental level at 0.927 MeV with the high B(E1)/B(E2) transition ratios and low normalized relative transfer strength is not reproduced, neither in the SQPM (see Table~\ref{BE_SQPM}) nor in the QPM \cite{LoI05}. Contrary to the IBM, two-octupole phonon states are shifted to higher energies $\sim 2.4 - 2.5$ MeV due to the Pauli exclusion principle. The lower-lying states (e.g. $0_4^{+}$ at 1.45 MeV) possess only small
two-octupole phonon admixtures. On the other hand, for the octupole-octupole isoscalar strength $\kappa^{(0)}_{30} = 7$ eV fm$^{-6}$, that reproduces the experimental energy of the first $1^-$ excited state, the SQPM predicts enhanced $E1$ transitions from the $K=0^-$ band to the g.s. band, e.g.
$B(E1;1_1^- \rightarrow 0_1^+)= 0.2$ e$^2$ fm$^2 = 0.08 $ W.u. As a consequence, even a small admixture of double octupole phonon configuration $[(30)_1(30)_1]$ in $0_3^{+}$
of about $0.3 - 0.6\%$ can account for the experimentally observed $B(E1)/B(E2)$ ratios (see Table~\ref{BE_SQPM}).

The SQPM and the QPM are quite accurate in nuclei with small ground-state correlations. Since in $^{232}$U the ground-state correlations (as tested for the SQPM) become
large, the effect of multi-phonon admixtures (three and more phonons) in the QPM that pushes two-phonon poles and consequently two-phonon
energies to lower values is then underestimated. In future QPM studies one also has to take into account the spin-quadrupole interaction that is known to influence the
density and structure of low-lying $0^+$ states \cite{Pya67}.

\begin{table*}
\caption{\label{QPM_structure} Phonon structure of the QPM $0^+$ states up to 2.6 MeV \cite{Sush15}.
The weights of the one-phonon $(\lambda\mu|_i)$ or the two-phonon
$[(\lambda\mu)_i(\lambda\mu)_i|]$ components are given in percent.  Only  main one-phonon
and two-phonon components are shown. Transfer factors $S(p,t)$ are normalized to 100 for the 0$^+_{g.s.}$.}
\begin{ruledtabular}
\begin{tabular}{cccc}\\
$K^{\pi}_n$ & $E_{n}(calc)$ & $S(p,t)_{calc}$ & Structure \\
\hline\\
0$_2^{+}$   &   0.51  &  25.81 & $(20)_1 91$ \\
0$_3^{+}$   &   1.33  &   4.09 & $(20)_2 90;[(22)_1(22)_1] 4$\\
0$_4^{+}$   &   1.45  &   4.01 & $(20)_3 27;(20)_5 27;[(22)_1(22)_1] 19;[(30)_1(30)_1] 2$ \\
0$_5^{+}$   &   1.71  &   0.03 & $(20)_3 57;(20)_5 23;(20)_4 12$    \\
0$_6^{+}$   &   1.79  &   0.18 & $(20)_4 79;[(22)_1(22)_1] 3$ \\
0$_7^{+}$   &   1.98  &   4.64 & $(20)_5 15;(20)_7 15;(20)_9 13;[(22)_1(22)_1] 16;[(30)_1(30)_1] 4$   \\
0$_8^{+}$   &   2.06  &   1.59 & $(20)_6 90$ \\
0$_9^{+}$   &   2.14  &   1.67 & $(20)_7 55;(20)_5 11;[(31)_1(31)_1] 3;[(32)_1(32)_1] 3$     \\
0$_{10}^{+}$   &   2.29  &   1.06 & $(20)_9 49;(20)_7 14;[(32)_1(32)_1] 11$ \\
0$_{11}^{+}$  & 2.30  &   0.06 & $(20)_9 4;[(44)_1(44)_1] 92$ \\
0$_{12}^{+}$  & 2.36  &   0.11 & $(20)_8 65;(20)_{12} 16;[(30)_1(30)_1] 2$      \\
0$_{13}^{+}$  & 2.43  &   0.59 & $(20)_{10} 7;[(32)_1(32)_1] 65$  \\
0$_{14}^{+}$  & 2.48  &   0.01 & $(20)_{10}63;(20)_9 11;[(32)_1(32)_1] 8$\\
0$_{15}^{+}$  & 2.51  &   1.73 & $(20)_{11} 43;[(30)_1(30)_1] 37$      \\
0$_{16}^{+}$  & 2.57  &   3.55 & $(20)_{11} 43;[(30)_1(30)_1] 29$      \\
\end{tabular}
\end{ruledtabular}
\end{table*}

\begin{table}
\caption{\label{BE_SQPM} Experimental and SQPM B(E1)/B(E2) transition ratios in
\(^{232}\)U between the states of the $0^{+}_3$ band and the $0^{-}_{1}$ and $0^{+}_{1}$ bands in units of $10^{-4}$ b$^{-1}$.}
\begin{ruledtabular}
\begin{tabular}{cccccc}\\
$K^{\pi}_{i}$ & J$_{i}$ & J$_{f1}$ & J$_{f2}$ & Exp. & SQPM \\
\hline \\
0$^{+}_3$ & $0^+$ & 1$^{-}_1$ & 2$^{+}_1$  &  44(7)  &  7.5 \\
          & $2^+$ & 1$^{-}_1$ & 0$^{+}_1$  & 150(30) &  15 \\
          & $2^+$ & 1$^{-}_1$ & 2$^{+}_1$  &  45(1)  &  11 \\
          & $2^+$ & 1$^{-}_1$ & 4$^{+}_1$  &  24(1)  &  5.8 \\
          & $2^+$ & 3$^{-}_1$ & 0$^{+}_1$  & 337(68) &  23 \\
          & $2^+$ & 3$^{-}_1$ & 2$^{+}_1$  & 101(3)  &  16 \\
          & $2^+$ & 3$^{-}_1$ & 4$^{+}_1$  &  54(1)  &  8.8 \\
\end{tabular}
\end{ruledtabular}
\end{table}

\begin{figure}
\begin{center}
\epsfig{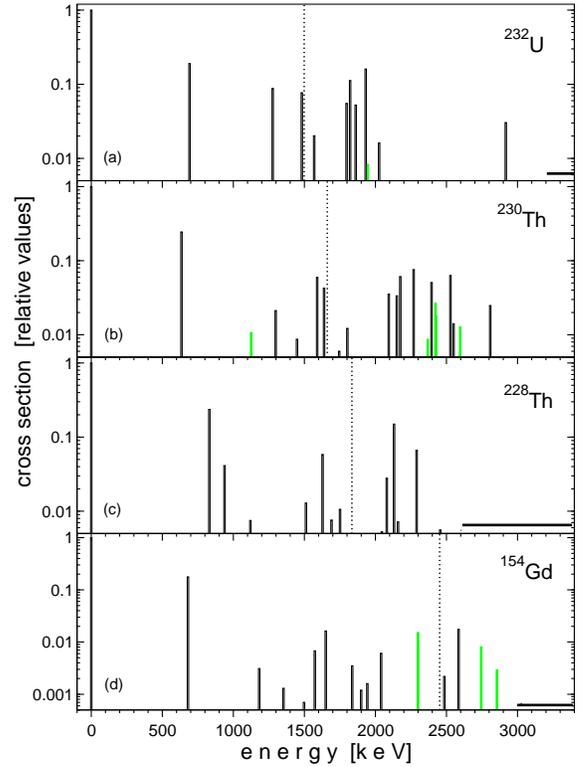} \caption{\label{fig:zero_all} (Color online)
Location and
the (p,t) strength of 0$^+$ states  in $^{232}$U, $^{230}$Th, $^{228}$Th
and $^{154}$Gd. The dotted lines indicate the pairing gap for each nucleus. Horizontal
lines indicate limitations in the investigation energy.}
\end{center}
\end{figure}

\subsection{\label{sec:den} To the density distribution  of excited 0$^+$ states }
As one can see in Fig.~\ref{fig:strength},  0$^+$ states are observed  in a limited area in
the form of a bump. Local groups of 2$^+$, 4$^+$   and 6$^+$ states  are shifted relative
to 0$^+$  states in the direction of higher energies.
The assumption that the $0^+$ states are localized mainly in a limited region, and that
the density of the 0$^+$ levels above 3 MeV is, at least, negligible was made in \cite{Lev09}.
With this purpose, the triton spectra from the $^{232}$Th(p,t)$^{230}$Th reaction
were measured for the energy range of 3 $\div$ 4 MeV, but only for the angles 12.5$^\circ$ and 26$^\circ$.
Two lines in the spectra meet the condition not only for the 0$^+$ state, but also for 6$^+$ states.
Fig.~\ref{fig:zero_all}  shows the 0$^+$ state
spectra  of studied actinides and  as an example one of the rare earth nucleus, $^{154}$Gd.
In the rare earth region, spin-parity values 0$^+$ were assigned for many nuclei using
the triton angular distributions for only three \cite{Mey06,Ili10} and even two \cite{Ber13} angles, exploring
the fact that the $L=0$ transfer angular distribution peaks strongly at forward angles.
As one could  see, some of the $L=2$ and 4 angular distributions also peak strongly
at forward angles. Therefore, some tentative assignments of spin 0$^+$ just below 3 MeV
(as for $^{154}$Gd) not actually belong to the 0$^+$ states. Only a detailed fitting  of
the angular distribution in a sufficiently large range of angles would allow to distinguish
between the 0$^+$ and 2$^+$ or 4$^+$  assignments.

\begin{figure}
\begin{center}
\epsfig{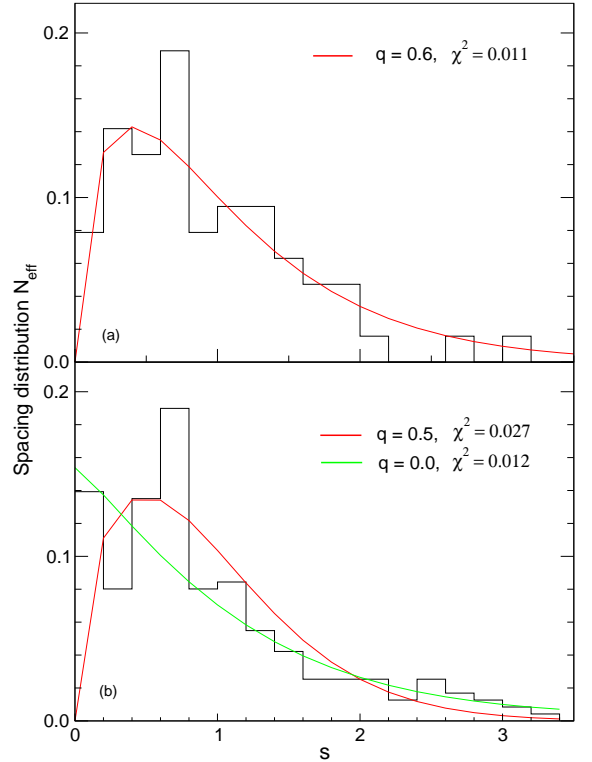}
\caption{\label{fig:stat_an} (Color online)
 Normalized
nearest-neighbor spacing as a function of a dimensionless spacing variable $s$
 and fit with the Brody distribution: (a) experimental data for $^{228,230}$Th, $^{232}$U
 and  $^{240}$Pu in the energy interval of $0 - 3$ MeV, (b)  calculated by
 the QPM data for $^{228,230}$Th, $^{232}$U
 in the energy interval of $0 - 4$ MeV.}
\end{center}
\end{figure}

At the same time, the IBM and the QPM predict  an increase in the number of 0$^+$ states with
increasing energy.  The impact of the inclusion of these additional levels can be seen
from the statistical analysis of the level density  for actinides, experimental in
the energy interval of $0 - 3$ MeV and predicted by the QPM  in the energy interval of
$0 - 4$ MeV (Fig.~\ref{fig:stat_an}). The Brody distribution was used
for fitting the normalized
nearest-neighbor spacing as a function of a dimensionless spacing variable $s$ \cite{Bro73}.
It was applied in \cite{Mey06} for  analysis of the 0$^+$ spectra in the rare earth
nuclei  testing for the ordered or chaotic (mixed) nature of these spectra.
The Brody distribution  describes systems with intermediate degrees of mixing depending
 on the parameter $q$, which ranges from 0 for a Poisson distribution (ordered nature) to 1 for a Wigner distribution (chaotic nature)
 \begin{equation}\label{eq:Brody}
N_{eff} = A s^q \exp{(-bs^{q+1})} \ ,
\end{equation}
where the parameters $b$  and $A$ are determined  by the value of $q$:
  $b = [\Gamma((2+q)/(1+q))]^{q+1}$
and $A=b(1+q)$. To get the value of  $\chi^2 $  parameter  $A$  was left free. In such a way,
the experimental data for $^{228,230}$Th \cite{Lev09,Lev13}, $^{232}$U and
$^{240}$Pu \cite{Spi13} are fitted by the Brody distribution for $q = 0.6$ (the same as for
the rare earth nuclei in \cite{Mey06})  with $\chi^2 = 0.011$.  The theoretical data
from \cite{LoI04,Sush15} can be fitted by the Brody distribution
for $q = 0.5$, but only with a worse $\chi^2 = 0.027$. In both cases, the obtained
values of the parameter $A$ are close to $A=b(1+q)$.
A much better fit is obtained for the Poisson distribution with $\chi^2 = 0.012$.
This means that the experimental 0$^+$ spectrum in the energy interval $0 - 3$ MeV is
intermediate between an ordered and chaotic nature, while the ordered nature is   preferred
for the theoretical spectrum in the energy interval $0-4$ MeV.
Besides that  the mean number of 0$^+$ states observed in one nucleus  in the energy range
of $0-3$ MeV is about 18, while the number of theoretically predicted 0$^+$ states
in the energy range of $0-4$ MeV is about 80.
Therefore, it is important to investigate at least the region $3 - 4$ MeV  for the presence
of additional 0$^+$ excitations.

The phenomenologic IBM-1 used in the present paper even
in its simplified two-parametric form is known for its capability to
study chaos and
transitions between order and chaos
in the properties of low-lying collective states of even-even nuclei
\cite{Alh90,Cej09}. In the microscopic QPM, an introduction of
multi-phonon states
(three and more) seems to be necessary to move from order towards chaos.
This idea is supported by the analysis performed for odd nuclei
\cite{Sto04,Sto13},
where the addition
of one-quasiparticle plus two-phonon states (i.e. $\rq$5-qp states$\rq$)
to the standard one-quasiparticle and one-quasiparticle plus one-phonon
states led to a fit
of the calculated $17/2^+$ $^{209}$Pb spectra to the Brody distribution
with the parameter $q=0.6$, thus corresponding to a transitional region
between order and chaos.


\section{Conclusion}

To summarize, in a high-resolution experiment the excited states of $^{232}$U have
been studied in the (p,t) transfer reaction.  162 levels were assigned, using a DWBA fit procedure.
Among them, 13 excited $0^+$ states have been found in this nucleus up to
an energy of 3.2 MeV, most of them have not been experimentally observed before.
Their accumulated strength makes up 84\,\% of the ground-state strength.
Firm assignments have been made for most of the 2$^+$, 4$^+$  and for about half of
the 6$^+$ states. These assignments allowed to identify the sequences of states,
which have the features of rotational bands with definite inertial parameters.
Moments of inertia are derived from these sequences.
Most of the values of the moments of inertia are not much higher than
 the value for the g.s. band. This indicates that they may correspond  mainly
 to a quadrupolar one-phonon structure of 0$^+$ states.

The experimental data have been compared to {\it spdf}-IBM and QPM calculations.
The IBM reproduces the main characteristics of the experimental transfer distribution,
namely the running sum of the (p,t) strengths and increased population of two groups
of 0\(^+\) excitations around 1 and 2 MeV, but the strength of the first excited $0^+$ state is
underestimated and the strength of the second $0^+$ state is overestimated.
Most of the calculated excitations have two \(pf\) bosons in their structure,
therefore being related to the presence of a double octupole structure.
Good agreement with experiment for the B(E1)/B(E2) transition ratios indicates also the
importance of the octupole degree of freedom.
The QPM reproduces the strong (p,t) strength of the first excited $0^+$
state due to its predicted pairing vibrational character
and lower (p,t) strengths for higher-lying $0^+$ states. It fails to
account for a rapid increase of the running sum of the (p,t) strength
above 1.8 MeV and predicts only minor double-octupole phonon components
in states below 2.4 MeV.
Both models fail to give a detailed explanation of the individual
states.

The comparison of the experimental nearest-neighbor spacing distribution
of the $0^+$ states in the region of $0-3$ MeV for four actinide
isotopes
($^{228,230}$Th, $^{232}$U and $^{240}$Pu) to the Brody distribution
revealed an intermediate character of the experimental $0^+$ spectrum
between order and chaos. A similar distribution for the data obtained from the QPM calculations
 in the region of $0-4$ MeV somewhat differs  from the experimental one
 and is closer to the ordered nature.  Though the increased role of
multi-phonon states in the model at higher energies means movement
to chaos. Therefore, (p,t) and (p,t$\gamma$) experiments for higher
energies could provide additional information on the nature
of $0^+$ excitations.

\section{Acknowledgements}

The work was supported by the DFG (C4-Gr894/2-3, Gu179/3, Jo391/2-3, Zl510/4-2);
by the National Programme for Sustainability I (2013-2020) by the state budget
of the Czech Republic, identification code: LO1406; by the Romanian Executive
Unit for Financing Higher Education, Research, Development and Innovation
under contracts Crt. 127/F4 and PN-II-ID-PCE-2011-3-0367.



\begin{thebibliography}{99}


\bibitem{Lev94}   A.~I.~Levon,  J.~de Boer, G.~Graw, R.~Hertenberger,
    D.~Hofer, J.~Kvasil, A.~L\"osch, E.~M\"uller-Zanotti, M.~W\"urkner, H.~Baltzer,
    V.~Grafen, and C.~G\"unther,  Nucl. Phys.  {\bf A576}, 267 (1994).

\bibitem{Wir04} H.-F.~Wirth, G.~Graw, S.~Christen, D.~Cutoiu, Y.~Eisermann,
    C.~G\"unther, R.~Hertenberger, J.~Jolie, A.~I.~Levon, O.~M\"oller, G.~Thiamova, P.~Thirolf,
    D.~Tonev, and N.~V.~Zamfir, Phys. Rev. C {\bf 69}, 044310 (2004).

\bibitem{Spi13}  M.~Spieker, D.~Bucurescu, J.~Endres, T.~Faestermann, R.~Hertenberger, S.~Pascu, S.~Skalacki,
    S.~Weber, H.-F.~Wirth, N.V.~Zamfir, and A.~Zilges, Phys. Rev. C {\bf 88}, 041303(R) (2013).

\bibitem{Les02} S.~R.~Lesher, A.~Aprahamian, L.~Trache, A.~Oros-Peusquens, S.~Deyliz, A.~Gollwitzer,
    R.~Hertenberger, B.~D.~Valnion, and G.~Graw, Phys. Rev. C {\bf 66}, 051305(R) (2002).

\bibitem{Mey06}  D.~A.~Meyer, V.~Wood, R.~F.~Casten, C.~R.~Fitzpatrick, G.~Graw, D.~Bucurescu, J.~Jolie,
    P.~von~Brentano, R.~Hertenberger, H.-F.~Wirth, N.~Braun, T.~Faestermann, S.~Heinze, J.~L.~Jerke,
    R.~Kr\"{u}cken, M.~Mahgoub, O.~M\"{o}ller, D.~M\"{u}cher, and C.~Scholl, Phys. Rev. C {\bf 74}, 044309 (2006).

\bibitem{Buc06}  D.~Bucurescu, G.~Graw, R.~Hertenberger, H.-F.~Wirth, N.~Lo~Iudice, A.~V.~Sushkov,
    N.~Yu.~Shirikova,  Y.~Sun, T.~Faestermann, R.~Kr\"{u}cken, M.~Mahgoub, J.~Jolie, P.~von~Brentano,
    N.~Braun, S.~Heinze, O.~M\"{o}ller, D.~M\"{u}cher, C.~Scholl, R.F.~Casten, D.A.~Meyer,
    Phys. Rev. C {\bf 73}, 064309 (2006).

\bibitem{Bet09}  L.~Bettermann, S.~Heinze, J.~Jolie, D.~M\"{u}cher, O.~M\"{o}ller, C.~Scholl, R.~F.~Casten,
    D.~A.~Meyer, G. Graw, R. Hertenberger, H.-F.~Wirth, and D.~Bucurescu, Phys. Rev. C {\bf 80}, 044333 (2009).

\bibitem{Ili10}  G.~Ilie, R.~F.~Casten, P.~von~Brentano, D.~Bucurescu, T.~Faestermann, G.~Graw, S.~Heinze,
    R.~Hertenberger, J.~Jolie,  R.~Kr\"{u}cken, D.~A.~Meyer, D.~M\"{u}cher, C.~Scholl, V.~Werner, R.~Winkler,
    and H.-F.~Wirth, Phys. Rev. C {\bf 82}, 024303 (2010).

 \bibitem{Ber13} C.~Bernards, R.~F.~Casten, V.~Werner, P.~von~Brentano, D.~Bucurescu, G.~Graw, S.~Heinze,
    R.~Hertenberger, J.~Jolie, S.~Lalkovski, D.~A.~Meyer, D.~M\"{u}cher, P.~Pejovic, C.~Scholl,
    and H.-F.~Wirth, Phys. Rev. C {\bf 87}, 024318 (2013).

\bibitem{Lev09} A.~I.~Levon, G.~Graw, Y.~Eisermann, R.~Hertenberger, J.~Jolie,
    N.~Yu.~Shirikova, A.~E.~Stuchbery, A.~V.~Sushkov, P.~G.~Thirolf, H.-F.~Wirth, and N.~V.~Zamfir,
    Phys. Rev. C {\bf 79}, 014318 (2009).

\bibitem{Lev13} A.~I.~Levon, G.~Graw, R.~Hertenberger, S.~Pascu, P.~G.~Thirolf,
    H.-F.~Wirth, and P.~Alexa,  Phys. Rev. C {\bf 88}, 014310 (2013).

\bibitem{LoI04} N.~Lo~Iudice, A.~V.~Sushkov, and N.~Yu.~Shirikova, Phys. Rev. C {\bf 70}, 064316 (2004).

\bibitem{LoI05} N.~Lo~Iudice, A.~V.~Sushkov, and N.~Yu.~Shirikova, Phys. Rev. C {\bf 72}, 034303 (2005).

\bibitem{Zam02} N.~V.~Zamfir, Jing-ye~Zhang, and R.~F.~Casten, Phys. Rev. C {\bf 66}, 057303 (2002).

\bibitem{Sun03} Yang~Sun, Ani~Aprahamian, Jing-ye~Zhang, and Ching-Tsai~Lee,
    Phys. Rev. C {\bf 68}, 061301(R) (2003).

\bibitem{She08} R.~K.~Sheline, P.~Alexa, Acta Phys. Pol. {\bf B 39}, 711 (2008).

\bibitem{Bro06} E.~Browne, Nucl. Data Sheets 107, 2579 (2006).

\bibitem{Zan91} E.~Zanotti, M.~Bisenberger, R.~Hertenberger, H.~Kader,
and  G.~Graw, Nucl. Instrum. Methods A {\bf 310}, 706 (1991).

\bibitem{Wir01} H.-F.~Wirth, Ph.~D.~thesis, Techn. Univ. M\"unchen, 2001,
(http://tumb1.biblio.tu-munchen.de/publ/diss/ph/2001/wirth.html).

\bibitem{Rie91} F.~Riess, Beschleunigerlaboratorium M\"unchen, Annual
                          Report, 1991, p.168.

 \bibitem{Bec69} F.~D.~Becchetti and G.~W.~Greenlees, Phys. Rev. {\bf 182},  1190 (1969).

\bibitem{Fly69} E.~R.~Flynn, D.~D.~Amstrong, J.~G.~Beery, and A.~G.~Blair,
                Phys. Rev. {\bf 182}, 1113 (1969).

\bibitem{Bec71} F.~D.~Becchetti and G.~W.~Greenlees, Proc. Third Int. Symp. on
               polarization phenomena in nuclear reactions, Medison, 1970,
               ed. H.~H.~Barshall and W.~Haeberli (University of Wisconsin
               Press, Medison, 1971) p.68

\bibitem{Kun} P.~D.~Kunz, computer code CHUCK3, University of Colorado,
unpublished.

\bibitem{Ard94} G.~Ardisson, M.~Hussonnois, J.~F.~LeDu, D.~Trubert, and C.~M.~Lederer,
    Phys. Rev. C {\bf 49}, 2963 (1994).

\bibitem{Wei72} R.~Weiss-Reuter, H.~M\"unzel, and G.~Pfennig, Phys. Rev. C {\bf 6}, 1425 (1972).


\bibitem{Iach87} F. Iachello and A. Arima, The Interacting Boson Model
(Cambridge University Press, Cambridge, England, 1987).

\bibitem{Eng87} J. Engel and F. Iachello, Nucl. Phys. {\bf A 472}, 61 (1987).

\bibitem{Zam01} N. V. Zamfir and D. Kusnezov, Phys. Rev. C {\bf 63}, 054306 (2001).

\bibitem{Zam03} N. V. Zamfir and D. Kusnezov, Phys. Rev. C {\bf 67}, 014305 (2003).

\bibitem{Rob12} L. M. Robledo and R. R. Rodriguez-Guzman,
                J. Phys. G: Nucl. Part. Phys. {\bf B 39}, 105013 (2012).

\bibitem{Rob13} L. M. Robledo and P. A. Butler, Phys. Rev. C {\bf 88}, 051302 (2013).

\bibitem{Naz84} W. Nazarewicz, P. Olanders, I. Ragnarsson, J. Dudek, G. A. Leander,
            P. Moller, and E. Ruchowska, Nucl. Phys. {\bf A 429}, 269 (1984).

\bibitem{But96} P. A. Butler and W. Nazarewicz, Rev. Mod. Phys., {\bf 68}, 349 (1996).


\bibitem{Cas88} R. F. Casten and D. D. Warner, Rev. Mod. Phys. {\bf 60}, 389 (1988).

\bibitem{Kuz90} D. Kusnezov, J. Phys. {\bf A 23}, 5673 (1990).

\bibitem{Kuz89} D. Kusnezov, J. Phys. {\bf A 22}, 4271 (1989).

\bibitem{Kuz_up} D. Kusnezov, computer code OCTUPOLE (unpublished).


\bibitem{Sol92} V.~G.~Soloviev,
{\it Theory of Atomic Nuclei: Quasiparticles and Phonons} (Institute of
Physics, Bristol, 1992).

\bibitem{Sol76} V.~G.~Soloviev, Theory of Complex Nuclei, Pergamon
Press, Oxford (1976).

\bibitem{Mol95} P.~M{\o}ller, J.~R.~Nix, W.~D.~Myers, and W.~J.~Swiatecki,
Atomic Data Nucl. Data Tables {\bf 59}, 185 (1995).

\bibitem{Mol97} P.~M{\o}ller, J.~R.~Nix, and K.-L.~Kratz,
Atomic Data Nucl. Data Tables {\bf 66}, 131 (1997).

\bibitem{Web98} T.~Weber, J.~de~Boer, K.~Freitag, J.~Gr\"oger,
C.~G\"unther, P.~Herzog,
V.~G.~Soloviev, A.~V.~Sushkov, and N.~Yu.~Shirikova,  Eur. Phys. J.
{\bf A3}, 25 (1998)

\bibitem{Sush15} A.V.~Sushkov (private communication).

\bibitem{Pya67} N.I. Pyatov, Ark. Phys. {\bf 36}, 667 (1967).

\bibitem{Bro73} T.~A.~Brody, Lett. Nuovo Cimento 7, 482 (1973).

\bibitem{Alh90} Y.~Alhassid, A.~Novoselsky, N.~Whelan, Phys. Rev. Lett.
{\bf 65}, 2971 (1990).

\bibitem{Cej09} P.~Cejnar and J.~Jolie, Progr. Part. Nucl. Phys. {\bf
62}, 62 (2009).

\bibitem{Sto04} Ch.~Stoyanov and V.~Zelevinsky, Phys. Rev. C{\bf 70},
014302 (2004).

\bibitem{Sto13} Ch.~Stoyanov, Rom. J. Phys. {\bf 58}, 1096 (2013).

\end{thebibliography}
\end{document}